\documentclass[aps,prd,showpacs,nofootinbib,twocolumn,showkeys, superscriptaddress,twocolumn,10pt]{revtex4-1}

\usepackage{graphicx}
\usepackage{amssymb}
\usepackage{amsmath}
\usepackage[svgnames]{xcolor}
\usepackage{mathtools,slashed}
\usepackage[retainorgcmds]{IEEEtrantools}
\usepackage{physics}
\usepackage[compat=1.1.0]{tikz-feynhand}
\usepackage{tikz}
\usepackage{epstopdf}
\usepackage[utf8]{inputenc}
\usepackage{url}
\usepackage[colorlinks,citecolor=DarkGreen,linkcolor=DarkRed,urlcolor=DarkBlue]{hyperref}

\usepackage{newtxtext}
\usepackage{newtxmath}
\usepackage[makeroom]{cancel}
\usepackage{epsfig}

\newcommand{\be}{\begin{IEEEeqnarray}{rCl}}
\newcommand{\ee}{\end{IEEEeqnarray}}
\usepackage{orcidlink}
\usepackage{hyperref}

\begin{document}

\title{Photon production from gluon splitting and fusion induced by a magnetic field of arbitrary strength in relativistic heavy-ion collisions}

\author{Alejandro Ayala\,\orcidlink{0000-0003-3929-9209}}
\affiliation{Instituto de Ciencias Nucleares, Universidad Nacional Aut\'onoma de M\'exico, Apartado Postal 70-543, Ciudad de México 04510,
  Mexico.}
\affiliation{Instituto de Física, Universidade de São Paulo, Rua do Matão, 1371, CEP 05508-090, São Paulo, SP, Brazil}
\affiliation{Instituto de Física Teórica, Universidade Estadual Paulista, Rua Dr. Bento Teobaldo Ferraz, 271 - Bloco II, 01140-070 São Paulo, SP, Brazil}
\author{Santiago Bernal-Langarica\,\orcidlink{0000-0003-0826-9349}}
\affiliation{Instituto de Ciencias Nucleares, Universidad Nacional Aut\'onoma de M\'exico, Apartado Postal 70-543, Ciudad de México 04510,
  Mexico.}
\author{Jos\'e Jorge Medina-Serna\,\orcidlink{0009-0002-2026-1542}}
\affiliation{Instituto de Ciencias Nucleares, Universidad Nacional Aut\'onoma de M\'exico, Apartado Postal 70-543, Ciudad de México 04510,
  Mexico.}
\author{Ana Julia Mizher\,\orcidlink{0000-0001-9502-9815}}
\affiliation{Instituto de Física, Universidade de São Paulo, Rua do Matão, 1371, CEP 05508-090, São Paulo, SP, Brazil}
\affiliation{Centro de Ciencias Exactas and Departamento de Ciencias Básicas, Facultad de Ciencias, Universidad del Bío-Bío, Casilla 447, Chillán, Chile}

\begin{abstract}

We compute the yield and elliptic flow coefficient for photons produced during the pre-equilibrium stage of semi-central relativistic heavy-ion collisions from the gluon fusion and splitting processes induced by the presence of a magnetic field. The calculation is performed for arbitrary values of the field strength. The pre-equilibrium gluon distribution is modeled with a glasma-inspired Bose-Einstein occupation factor, as well as with an anisotropic distribution that accounts for the rapid initial expansion along the beam axis. In both cases, we find a very good agreement between the calculation and the experimental data from PHENIX. Most notably, the shape and strength of the elliptic flow coefficient are described quite well, representing a step towards solving the outstanding photon puzzle. 

\end{abstract}

\maketitle

\section{Introduction}\label{sec:I}

The study of strongly interacting matter in the presence of external fields has attracted considerable attention over the past decades, as it provides access to aspects of the underlying interactions that are otherwise difficult to probe~\cite{Gusynin:1999pq,Mueller:2014tea,Avancini:2015ady,Avancini:2016fgq,Avancini:2018svs,Carlomagno:2022inu,Ayala:2018zat,Avancini:2021pmi,Coppola:2018vkw,Coppola:2023mmq,Li:2016dta,Cancino:2026xpk,Ayala:2019akk,Ayala:2021lor,Ayala:2020muk,Ayala:2020dxs,Rojas:2008sg,Castano-Yepes:2022luw,Castano-Yepes:2023brq,Castano-Yepes:2024ltr,Moreira:2022dwo,Castano-Yepes:2024ctr,Wang:2026xsm,Ayala:2025qag,Ayala:2026eja}. In particular, magnetic fields constitute a powerful and versatile tool across a wide range of physical systems, including condensed matter, astrophysics, and nuclear physics~\cite{Hattori:2023egw,Adhikari:2024bfa,Sinha:2026gkk,Lai:2000at,ManrezaParet:2020poe,Ayala:2021nhx,Bruckmann:2017pft,Ayala:2006sv}. In particular, it has been suggested that in semi-central relativistic heavy-ion collisions, strong magnetic fields are produced~\cite{Skokov:2009qp}, with rapidly fading intensities~\cite{Yan:2021zjc}. The peak intensities can be as high as $|eB| \sim 10^{19}$~G for RHIC energies~\cite{STAR:2019wlg, Brandenburg:2021lnj}. Because of these possibly large magnitudes, the effects of magnetic fields in the initial stages of the collision have to be taken into account in order to provide a complete description of the evolution of such systems~\cite{Sahu:2025tmb,Ayala:2025jpr,Ayala:2024jvc}. In this context, the excess yield~\cite{PHENIX:2022rsx} and the strength of the elliptic flow coefficient $v_2$ of direct photons, which is comparable to that of hadrons~\cite{PHENIX:2011oxq}, collectively referred to as the \textit{direct photon puzzle}, have been associated with magnetic field-induced processes~\cite{Tuchin:2010gx,Tuchin:2013ie,Tuchin:2012mf,Basar:2012bp}. For low transverse momentum, the yield is thought to be dominated by thermal photons originating from the very early stages of the collision thermal history. This early emission of direct photons appears to be confirmed by the dependence of $v_2$ on the photon transverse momentum, which, for large values, is consistent with zero. Since photons are colorless penetrating probes, they can only be boosted at their production times. Therefore, they should originate from the early stages, where expansion velocities are relatively small. To address this puzzle, several approaches have been proposed, including hydrodynamical scenarios~\cite{Gale:2021emg, Niemi:2015qia, Chatterjee:2011dw, Dasgupta:2019whr}, radiation emitted during the pre-equilibrium stage~\cite{Monnai:2014kqa, Berges:2017eom, Oliva:2017pri, Garcia-Montero:2023lrd}, transport models~\cite{Xu:2004mz, Kasmaei:2019ofu}, intermediate semi-QGP states~\cite{Gale:2014dfa, Lee:2014pwa}, and others, as reviewed in Refs.~\cite{David:2019wpt, Blau:2023bvi}. 

Another alternative scenario arises from the presence of magnetic fields, which naturally induce the emission of anisotropic electromagnetic radiation and contribute to the photon positive $v_2$ without requiring a linkage of its strength to the flow properties of the system. Among the possible channels opened by the magnetic field for photon production are the induced bremsstrahlung and pair annihilation in the QGP~\cite{Tuchin:2014iua, Wang:2020dsr, Wang:2022jxx},  electromagnetic radiation induced by the QED $ \times $ QCD conformal anomaly~\cite{Basar:2012bp}, photons originating from synchrotron radiation from a rotating and magnetized QGP~\cite{Buzzegoli:2026sji}, and photons radiated from $2\to 2$ scattering processes among quarks and gluons immersed in weak magnetic fields~\cite{Sun:2023pil, Sun:2024vsz}. Holographic methods to describe photon production from a strongly coupled plasma in the presence of a magnetic field have also been employed~\cite{Muller:2013ila, Arciniega:2013dqa, Avila:2022cpa}. Recently, photon production due to stochastic fluctuations of magnetic fields has also been explored~\cite{Castano-Yepes:2024vlj}, as well as the thermal photon emission from a magnetized plasma~\cite{Ouyang:2026bhn}. In a series of recent works, we have concentrated on describing two other possible channels for photon production in a magnetic field: gluon fusion and splitting during pre-equilibrium~\cite{Ayala:2017vex, Ayala:2019jey, Ayala:2022zhu, Ayala:2024ucr}. 

Recall that during pre-equilibrium, the gluon occupation number is enhanced with respect to that of quarks by a factor of $\alpha_s ^2$~\cite{Baier:2000sb, Garcia-Montero:2019vju, Monnai:2019vup}, therefore, although the amplitudes for gluon fusion and splitting are suppressed with respect to the leading order in the perturbative expansion, meaning quark or anti-quark splitting and quark -- anti-quark annihilation, the gluon contributions are enhanced relative to the quark contributions. Furthermore, by taking into account Pauli blocking in the final state, processes involving quarks in the initial or final state are further suppressed. For these reasons, gluon fusion and splitting become important channels for the production of photons during the pre-equilibrium stage. These channels were studied in Refs.~\cite{Ayala:2017vex, Ayala:2019jey} under the strong magnetic field approximation, which is only valid for the low transverse momentum region of the spectrum, and using a vertex function~\cite{Ayala:2022zhu} that did not take into account the complete tensor structure and the symmetry properties expected for these processes. The complete tensor structure of the effective vertex was studied from first principles in Ref.~\cite{Ayala:2024ucr}, and the one-loop level two-gluon one-photon vertex was computed in the intermediate field regime. Nonetheless, the contribution to the photon yield and $v_2$ arising from this vertex was not computed. The same channels were studied on an event-by-event basis during the early stage of high energy nuclear collisions, using the initial gluon distribution from the McLerran-Venugopalan model~\cite{Jia:2022awu}.

In this work, we undertake this task and compute the contributions to the photon yield and elliptic flow coefficient from gluon fusion and splitting, considering magnetic fields of arbitrary strength. The work is organized as follows: for the completeness of the discussion, in Sec.~\ref{sec:II} we summarize the findings of Ref.~\cite{Ayala:2024ucr} for the general structure of the two-gluon one-photon vertex in the presence of a constant magnetic field. In Sec.~\ref{sec:III}, we summarize the computation of the two-gluon one-photon effective vertex at one-loop level and give its explicit expression without requiring any approximation on the hierarchy of energy scales. In Sec.~\ref{sec:IV}, we present the results for the photon yield and $v_2$ produced by gluon fusion and splitting for arbitrary field strengths. Finally, in Sec.~\ref{sec:V} we summarize and present our conclusions. In the appendices, we list the expressions that appear in the computation of the traces and the vertex.

\section{General structure of the two-gluon one-photon vertex}\label{sec:II}

\begin{figure}
\centering
\begin{tikzpicture}
\begin{feynhand}
\setlength{\feynhandblobsize}{15mm}
\vertex (a) at (-2,2) {$a, \;\mu$}; \vertex  (b) at (-2,-2) {$b,\;\nu$};  \vertex[NEblob] (c) at (0,0) {B}; \vertex (d) at (2.5,0) {$\alpha$};
\propag [glu] (a) to [edge label' =$p_1$] (c);
\propag [glu] (b) to [edge label' =$p_2$] (c) ;
\propag [bos] (c) to [edge label =$q$] (d);
%\propag [sca] (c) to [quarter left, edge label=$K$] (d);
\end{feynhand}
\end{tikzpicture}
\caption{General representation of the two-gluon one-photon vertex. The shaded blob represents the effect of a magnetic field.}
\label{fig:vertex}
\end{figure}

The effective vertex describing the interaction between two gluons and one photon is depicted in Fig.~\ref{fig:vertex}. This effective vertex was studied in detail in Ref.~\cite{Ayala:2024ucr} and here we present the most important steps that were followed in deriving its general tensor structure. This vertex is a third-rank tensor, which can be denoted as $\Gamma_{ab}^{\mu\nu\alpha} (p_1, p_2, q)$, where $p_1$, $p_2$ are the gluons momenta with Lorentz indices $\mu$ and $\nu$, $q$ is the photon momentum with Lorentz index $\alpha$, $a$ and $b$ are the gluons color indices. Recall that, in the absence of a magnetic field, this vertex must vanish as prescribed by Furry's theorem.

The vertex is restricted by its symmetry properties: gauge invariance implies that it must be transverse when contracted with each of the external gauge boson momenta. The indistinguishability of gluons requires the vertex to be symmetric under the exchange of gluon indices and momenta. Since charge ($C$) and parity ($P$) are good quantum numbers for the strong and electromagnetic interactions, and there are three neutral gauge bosons involved as external states, the vertex must be invariant under the combined action of the $CP$ transformation.

To find the correct tensor structure of the vertex, the previously stated symmetry properties must be elucidated into the properties of a chosen tensor basis and its coefficients that comprise the vertex. To this end, we begin with the photon leg and introduce a set of four linearly independent vectors that span the four-dimensional Minkowski space-time and take into account the presence of the magnetic field; that is, we take the photon momentum and three general polarization vectors. These vectors can be chosen as the Ritus basis~\cite{Papanyan:1972cv, Papanyan:1974xa}, given by
\be
q^\mu &, &\nonumber\\
l_q ^\mu & \equiv & \hat{F}^{\mu\nu} q_\nu, \nonumber \\
l_q ^{*\mu} & \equiv & \hat{F}^{*\mu\nu} q_\nu, \nonumber \\
k_q ^\mu & \equiv & \frac{q^2}{l_q ^2} \hat{F}^{\mu \nu} \hat{F}_{\nu\beta} q^\beta + q^\mu ,
\ee
where 
\begin{equation}
    \hat{F}^{\mu\nu} = \flatfrac{F^{\mu\nu}}{|B|},
\end{equation}
with $F^{\mu\nu}$ the electromagnetic field strength tensor, $F^{*\mu\nu}$ its dual, and $|B|$ the magnetic field intensity. The three polarization vectors are transverse with respect to the photon momentum, form an orthogonal basis, and satisfy the closure relation
\begin{equation}
    g^{\mu\nu} = \frac{q^\mu q^\nu}{q^2} + \frac{l_q ^\mu l_q ^\nu}{l_q ^2} + \frac{l_q ^{*\mu} l_q ^{*\nu}}{l_q ^{*2}} + \frac{k_q ^\mu k_q ^\nu}{k_q ^2}.
\end{equation}
Notice that, for on-shell photons, $q^2 = 0$, and hence, $k_q ^\mu = q^\mu$, which is consistent with the fact that, for real photons, there are only two physical polarizations: $l_q ^\mu$ and $l_q ^{*\mu}$.

For a constant and homogeneous magnetic field in the $\hat{z}$ direction, the polarization vectors of the Ritus base are normalized to 
\be 
l_q ^\mu & = & \frac{1}{\sqrt{-q_\perp ^2}} \left(0, q_y, -q_x, 0\right), \nonumber \\
l_q ^{*\mu} & = & \frac{1}{\sqrt{q_\parallel ^2}} \left( q_z, 0, 0, \omega_q \right), \nonumber \\
k_q ^\mu & = & \frac{1}{\sqrt{q^2 q_\perp^2 q_\parallel^2}} \left( q_\perp^2 \omega_q, -q_\parallel^2 q_y, q_z  \right),
\ee
where $q_\perp^2 = -(q_x^2 + q_y ^2)$ and $q_\parallel^2 = (\omega_q^2 - q_z^2)$. This coincides with the polarization vectors discussed in Ref.~\cite{Hattori:2017xoo}.

Now, we can repeat this procedure for each of the gluons legs, in such a manner that we obtain their corresponding polarization vectors, namely
\be
\text{photon } \alpha & \to & q^\alpha, \; l_q ^\alpha, \; l_q ^{*\alpha}, \; k_q ^\alpha \nonumber \\
\text{gluon } \mu,\; a & \to & p_{1 a}^\mu, \; l_{p_1 a} ^\mu, \; l_{p_1 a} ^{*\mu}, \; k_{p_1 a} ^\mu \nonumber \\
\text{gluon } \nu,\; b & \to & p_{2 b}^\nu, \; l_{p_2 b} ^\nu, \; l_{p_2 b} ^{*\nu}, \; k_{p_2 b} ^\nu.
\ee
In this way, the $i-$th element of the basis is given by one of the products of the three polarization vectors, where each factor is taken from one of the sets of polarization vectors for each particle
\be
    \Gamma_{ab\;i}^{\mu\nu\alpha} & \in & \left\lbrace \left[l_{p_1 a} ^\mu, l_{p_1 a} ^{*\mu}, k_{p_1 a}^\mu \right] \otimes \left[l_{p_2 b} ^\nu, l_{p_2 b} ^{*\nu}, k_{p_2 b}^\nu \right] \right. \nonumber \\
    & & \left. \; \otimes \left[l_{q} ^\alpha, l_{q} ^{*\alpha}, k_{q}^\alpha \right] \right\rbrace,
\ee
which gives 27 different tensor structures in total, which must be restricted by the symmetry properties previously discussed. By means of the symmetry under gluon exchange, the number of tensor structures is reduced to 18.

On the other hand, if we consider the transformation properties of each remaining structure under $C$ and $P$, and impose that the vertex must be invariant under the combined action of $CP$, then the scalar coefficients that go with each tensor structure can be classified according to their properties under $C$ and $P$. In terms of any of the momentum vectors of the external bosons, $p_m ^\mu$, with $m = 1,2,3$, then the available Lorentz scalars take the form
\be
S_{1mn}^{++} & = & \left(p_m \cdot p_n\right),\nonumber\\
S_{2mn}^{++} & = & \left(p_m \cdot p_n\right)_\perp, \nonumber\\
S_{3mn} ^{-+} & = & p_m ^\mu \hat{F}_{\mu\nu}p_n ^\nu, \nonumber \\
S_{4mn} ^{--} & =& p_m ^\mu \hat{F}_{\mu\nu} ^* p_n ^\nu,
\ee
where we denoted $S_{jmn}^{CP}$ to show its properties under $C$ and $P$.

If now we restrict ourselves to on-shell gauge bosons, then as it was previously stated, the $k_p$ polarization vector reduces to the particle momentum; therefore, all the terms that include it are zero, and the total number of tensor structures reduces further to 5. Even more, for on-shell gauge bosons, the conservation of energy-momentum,
\be
\omega_{p_1} + \omega_{p_2} & = & \omega_q ,\nonumber \\
\mathbf{p}_1 + \mathbf{p}_2  & = & \mathbf{q}, 
\ee
implies that the gauge bosons must be collinear; hence, the gluons momenta can be written in terms of the photon momentum. Additionally, only the Lorentz scalar $S_{2mn}^{++}$ is non-zero; therefore, all tensor structures whose transformation properties under $C$ or $P$ are odd are not available to express the on-shell vertex. Therefore, by imposing all the symmetry properties, we arrive at
\be
\label{eq:TenStruct}
\Gamma_{ab} ^{\mu\nu\alpha} (p_1, p_2, q)_\text{on-shell} & = & a_1 ^{++} l_{p_1 a}^\mu l_{p_2 b}^\nu l_{q}^\alpha + a_2 ^{++} l_{p_1 a}^{*\mu} l_{p_2 b}^{*\nu} l_{q}^\alpha \nonumber\\
& & + \: \frac{a_{10} ^{++}}{\sqrt{2}} \left( l_{p_1 a}^{\mu} l_{p_2 b}^{*\nu} + l_{p_1 a}^{*\mu} l_{p_2 b}^{\nu} \right) l_{q}^{*\alpha}. \IEEEeqnarraynumspace
\ee
From this result, it can be seen that only 3 independent tensor structures are needed to describe the two-gluon one-photon effective vertex in the presence of a magnetic field, for on-shell gauge bosons. Additionally, since the basis is orthonormal, it is easy to obtain the coefficients $a_i ^{++}$ by projecting the vertex $\Gamma_{ab} ^{\mu\nu\alpha}$ onto each of the tensor structures available in the basis.

\section{One-loop approximation for the two-gluon one-photon for magnetic fields of arbitrary strength}\label{sec:III}

\begin{figure}
\begin{center}
\begin{tikzpicture}
\begin{feynhand}
\vertex (a) at (0,0.75); \vertex  (b) at (0,-0.75);  \vertex (c) at (2,0.75); \vertex (d) at (2,-0.75); \vertex (e) at (3.5,0); \vertex (f) at (5,0); 
\vertex  (g) at (2,1) {$\mathcal{A}$};
\propag [glu] (a) to [edge label' =$p_1$] (c);
\propag [glu] (b) to [edge label' =$p_2$] (d) ;
\propag [fer] (d) to  (c);
\propag [fer] (c) to  (e);
\propag [fer] (e) to  (d);
\propag [bos] (e) to [edge label =$q$] (f);
%\propag [sca] (c) to [quarter left, edge label=$K$] (d);
\end{feynhand}
\end{tikzpicture}
\end{center}
\begin{center}
\begin{tikzpicture}
\begin{feynhand}
\vertex (a) at (0,0.75); \vertex  (b) at (0,-0.75) {};  \vertex  (c) at (2,0.75) ;
\vertex  (g) at (2,1) {$\mathcal{B}$}; \vertex (d) at (2,-0.75); \vertex (e) at (3.5,0); \vertex (f) at (5,0); 
\propag [glu] (a) to [edge label' =$p_1$] (c);
\propag [glu] (b) to [edge label' =$p_2$] (d) ;
\propag [antfer] (d) to  (c);
\propag [antfer] (c) to  (e);
\propag [antfer] (e) to  (d);
\propag [bos] (e) to [edge label =$q$] (f);
%\propag [sca] (c) to [quarter left, edge label=$K$] (d);
\end{feynhand}
\end{tikzpicture}
\end{center}
\caption{Diagrams representing the one-loop order contribution to the two-gluon one-photon vertex. Diagram $\mathcal{B}$ represents the charge conjugate of diagram $\mathcal{A}$. The four-momentum vectors are chosen such that $q=p_1+p_2$.}
\label{fig:Fusion}
\end{figure}

The leading order contribution to the vertex, in terms of the couplings $\alpha_s$ and $\alpha_{em}$, which describes gluon fusion and splitting, is shown in Fig.~\ref{fig:Fusion}. The amplitude corresponds to the sum of both Feynman diagrams represented as two fermion triangles with two gluons and one photon attached to the vertices, with the charge in one diagram and the charge in the other flowing in opposite directions, where the internal fermion propagator is considered in the presence of the magnetic field. This amplitude was studied in detail in Ref.~\cite{Ayala:2024ucr}, and here we present a summary of the calculation and refer the reader to that reference for a thorough discussion. 

To start, recall that the magnetic field breaks Lorentz invariance, and as a consequence, each diagram must be computed starting from configuration space. The translationally invariant part of the fermion propagator can be written in the Schwinger proper time representation as
\be
S(p) & = & \int_0 ^\infty \frac{\dd{s}}{\cos(q_f B s)} e^{is\left( p_\parallel^2 + p_\perp ^2 \frac{\tan(q_f B s)}{q_f B s} - m_f ^2 + i\epsilon \right)} \nonumber \\
& & \times \left[ e^{iq_f B s \Sigma_3} \left( m_f + \slashed{p}_\parallel \right) + \frac{\slashed{p}_\perp}{\cos(q_f B s)} \right],
\ee
where $m_f$ is the mass of the quark with flavor $f$ and $\Sigma_3 = i\gamma_1 \gamma_2$. The explicit sum of the diagrams can be written as
\be
\Gamma_{ab} ^{\mu\nu\alpha} & = & -i g_s^2 q_f \int \dd[4]{x} \dd[4]{y} \dd[4]{z} \int \frac{\dd[4]{r_1}}{(2\pi)^4} \frac{\dd[4]{r_2}}{(2\pi)^4} \frac{\dd[4]{r_3}}{(2\pi)^4} \nonumber \\
& & \times e^{ir_3 \cdot (y-x)} e^{ir_2 \cdot (x-z)} e^{ir_1 \cdot (z-y)} e^{-i p_1 \cdot z} e^{-i p_2 \cdot y} e^{i q\cdot x} \nonumber \\
& & \times \: \left\lbrace \Tr[\gamma^\alpha S(r_2) \gamma^\mu t_a S(r_1) \gamma^\nu t_b S(r_3) ]\Phi(x,y,z,x) \right. \nonumber \\
& & \left. +\: \Tr[\gamma^\alpha S(r_3) \gamma^\nu t_b S(r_1) \gamma^\mu t_a S(r_2) ]\Phi(x,z,y,x) \right\rbrace,\nonumber\\ \IEEEeqnarraynumspace
\ee
where 
\be
\Phi(x,y,z,x) & \equiv & \Phi(x,y) \Phi(y,z) \Phi(z,x), \nonumber \\
\Phi(x,z,y,x) & \equiv & \Phi(x,z) \Phi(z,y) \Phi(y,x) = \Phi^* (x,y,z,x),\IEEEeqnarraynumspace
\ee
are the product of the Schwinger phases, which is a gauge invariant quantity~\cite{Ayala:2020muk}, defined as
\begin{equation}
    \Phi(x,y) = \exp[iq_f \int_y ^x \dd{\xi^\mu} \left[ A_\mu + \frac{1}{2}F_{\mu\nu}\left( \xi - y \right)^\nu \right] ],
\end{equation}
with $q_f$ being the charge of quark with flavor $f$, $g$ the quark gluon coupling, $t_c = \flatfrac{\lambda}{2}$, with $\lambda_c$ the Gell-Mann matrices. In order to describe a constant magnetic field in the $\hat{z}$ direction, we write the vector potential in the symmetric gauge, $A^\mu = \tfrac{B}{2}(0,-y,x,0)$.

The integration over the space-time variables gives rise to the overall delta function signaling the energy-momentum conservation. We introduce the definitions
\be
c_j & \equiv & \cos(q_f B s_j), \nonumber \\
t_j & \equiv & \tan(q_f B s_j), \nonumber \\
e_j & \equiv & c_j e^{i \text{sign}(q_f B) q_f B \Sigma_3},
\ee
where $s_j$ are the Schwinger proper-time parameters associated with the propagator of each of the internal fermion lines.

We first discuss photon emission by the gluon fusion process. After a long but straightforward computation of Gaussian integrals over the loop momenta and by taking the on-shell limit, we arrive at 

\begin{widetext}
\be 
\Gamma_{ab}^{\mu\nu\alpha} & = & -i \frac{g_s^2 q_f^2 B}{(2\pi)^2} \Tr[t_a t_b] \delta^{(4)}(p_1 + p_2 - q) 
\int_0 ^\infty \frac{\dd{s_1} \dd{s_2} \dd{s_3}}{c_1^2 c_2 ^2 c_3 ^2} \left(\frac{1}{t_1 t_2 t_3 - t_1 - t_2 - t_3}\right) \left( \frac{e^{-i s m_f ^2}}{s} \right) \nonumber\\
& & \times\: e^{-\frac{i}{s} \left(s_1 s_3 \omega_{p_1} ^2 + s_2 s_3 \omega_{p_2}^2 + s_1 s_2 \omega_q ^2\right) \frac{q_\perp ^2}{\omega_q ^2}} e^{-\frac{i}{\omega_q ^2} \frac{q_\perp ^2}{|q_f B|} \left(\frac{t_1 t_3 \omega_{p_1}^2 + t_2 t_3 \omega_{p_2}^2 + t_1 t_2 \omega_q ^2}{t_1 t_2 t_3 - t_1 - t_2 - t_3}\right)} \sum_{j = 1} ^{13} \left( \mathcal{T}_{\mathcal{A} j} ^{\mu\nu\alpha} + \mathcal{T}_{\mathcal{B} j} ^{\mu\nu\alpha} \right),
\label{vertix}
\ee
where $s_1 + s_2 + s_3 = s$ and the traces over Dirac space corresponding to diagrams $\mathcal{A}$ and $\mathcal{B}$ of Fig.~\ref{fig:Fusion} are given in Appendix~\ref{Appendix:traces}. This expression is an exact one-loop result and we proceed to work it out without making any approximation. This is in contrast to the calculation done in Ref.~\cite{Ayala:2024ucr} where only the intermediate field strength regime was considered. To this end, we can make the change of variable $s_i \equiv v_i s$, for $i = 1, 2, 3$ with $v_1 + v_2 + v_3 = 1$, and thus $v_3 = 1 - v_2 - v_1$. Therefore, the remaining integrals to be performed are over the variables $\tau$, $v_1$ and $v_2$ within the domains $0\leq \tau < \infty$, $0\leq v_1 \leq 1$ and $0 \leq v_2 \leq 1 - v_1$, where we have also used the scaled variable $\tau \equiv \flatfrac{i s}{q_f B}$. 

From Eq.~(\ref{vertix}) we can compute the coefficients of the tensor basis,  Eq.~\eqref{eq:TenStruct}. These coefficients can be written as integrals of the form
\be
    a_i^{++} & \!=\! & -\frac{i g_s^2 \Tr[t_a t_b]}{8\pi^2 B}\int_0 ^\infty \!\!\dd{\tau} \int_0 ^{1} \dd{v_1} \int_0 ^{1-v_1} \!\!\!\!\!\dd{v_2} e^{-\frac{m_f^2}{q_f B} \tau} e^{\sin^2 \theta \mathcal{F} (\tau, v_1, v_2)} \sin \theta \csch^2\tau \left[ \mathcal{G}_i (\tau, v_1, v_2) + \frac{\tau}{2}\mathcal{H}_i (\tau, v_1, v_2) \sin^2 \theta \right], \IEEEeqnarraynumspace
\ee
where $\theta$ is the angle between the photon momentum and the magnetic field. $\mathcal{F} $, $\mathcal{G}_i$ and $\mathcal{H}_i$ are scalar functions given in Appendix~\ref{Appendix:functions}. The remaining integrals can be numerically evaluated using a Monte Carlo VEGAS routine~\cite{Galassi:2019czg}.

Since the basis is orthonormal and the cross-section for the process of interest is written in terms of the square modulus of the effective vertex, the squared matrix element is expressed as
\begin{equation}
    |\Gamma|^2 = {|a_1 ^{++}}|^2 + |{a_2 ^{++}}|^2 + |{a_{10} ^{++}}|^2,
\end{equation}
for each quark flavor. 

Figure~\ref{fig:NumVertex} shows the behavior of the squared matrix element as a function of the photon energy $\omega_q$, the angle $\theta$ between the photon and the magnetic field directions and the magnetic field strength scaled by the squared of the pion mass $|eB|/m_\pi^2$, for $\alpha_s = 0.3$ and summed over the quark flavors $f = u,d,s$, where the mass of the quarks was considered as $m_u = 5$ MeV, $m_d = 10$ MeV and $m_s = 100$ MeV. In particular, notice that the matrix element, as a function of the photon energy, reaches a maximum to then slowly decrease, while as a function of the angle with respect to the magnetic field, it only has a maximum for $\theta = \flatfrac{\pi}{2}$ for low energy configurations, signaling that the preferred emission transverse to the magnetic field comes from photons with small energy. The matrix element is symmetric around $\theta=\pi$ and vanishes at $\theta=0,\ 2\pi$. Also, the matrix element, as a function of the magnetic field strength rapidly grows and reaches a plateau. This regime happens sooner as a function of the field strength for the lower photon energies.
\begin{figure}[t!]
    \centering
    \includegraphics[width=0.33\textwidth]{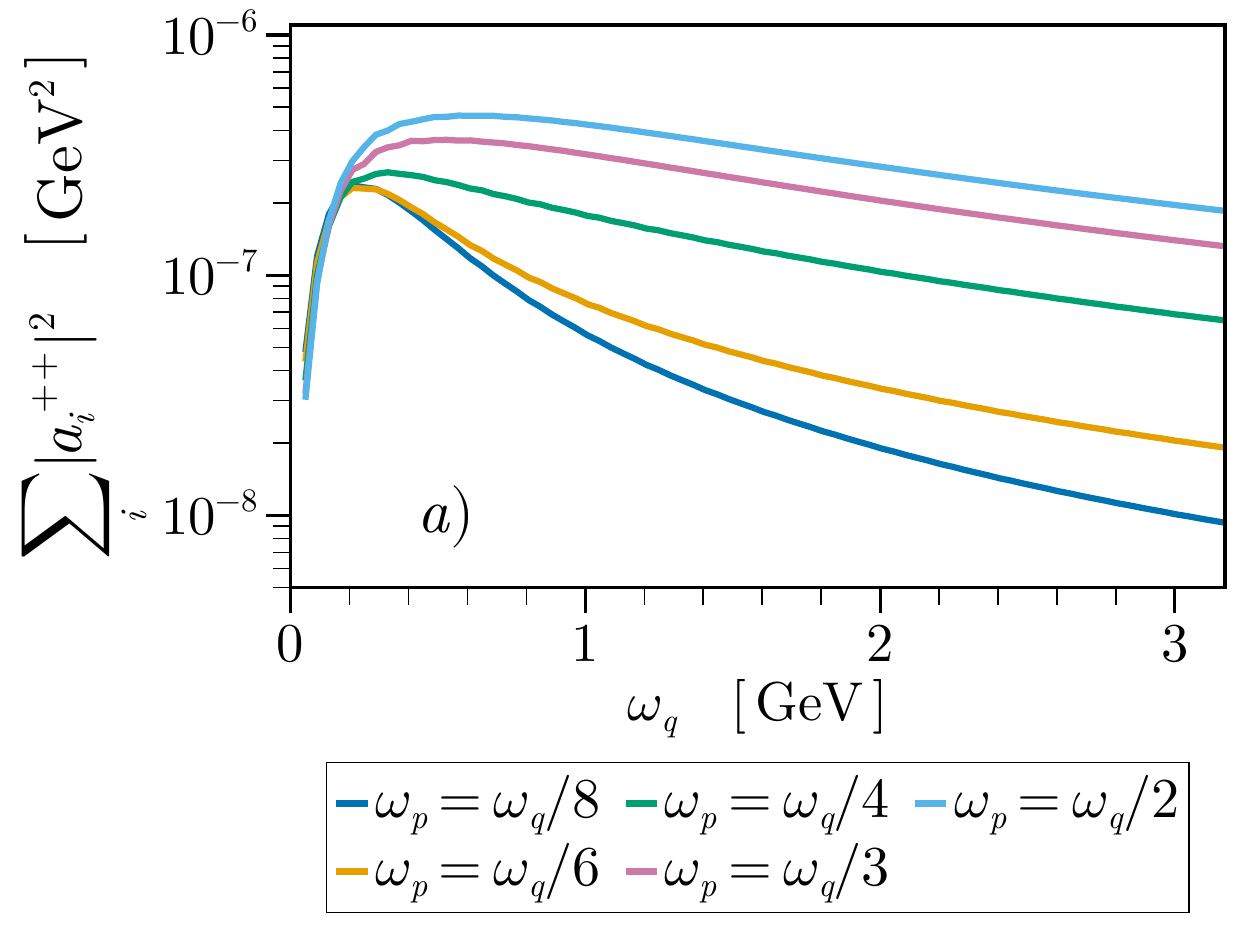} \hfill
    \includegraphics[width=0.33\textwidth]{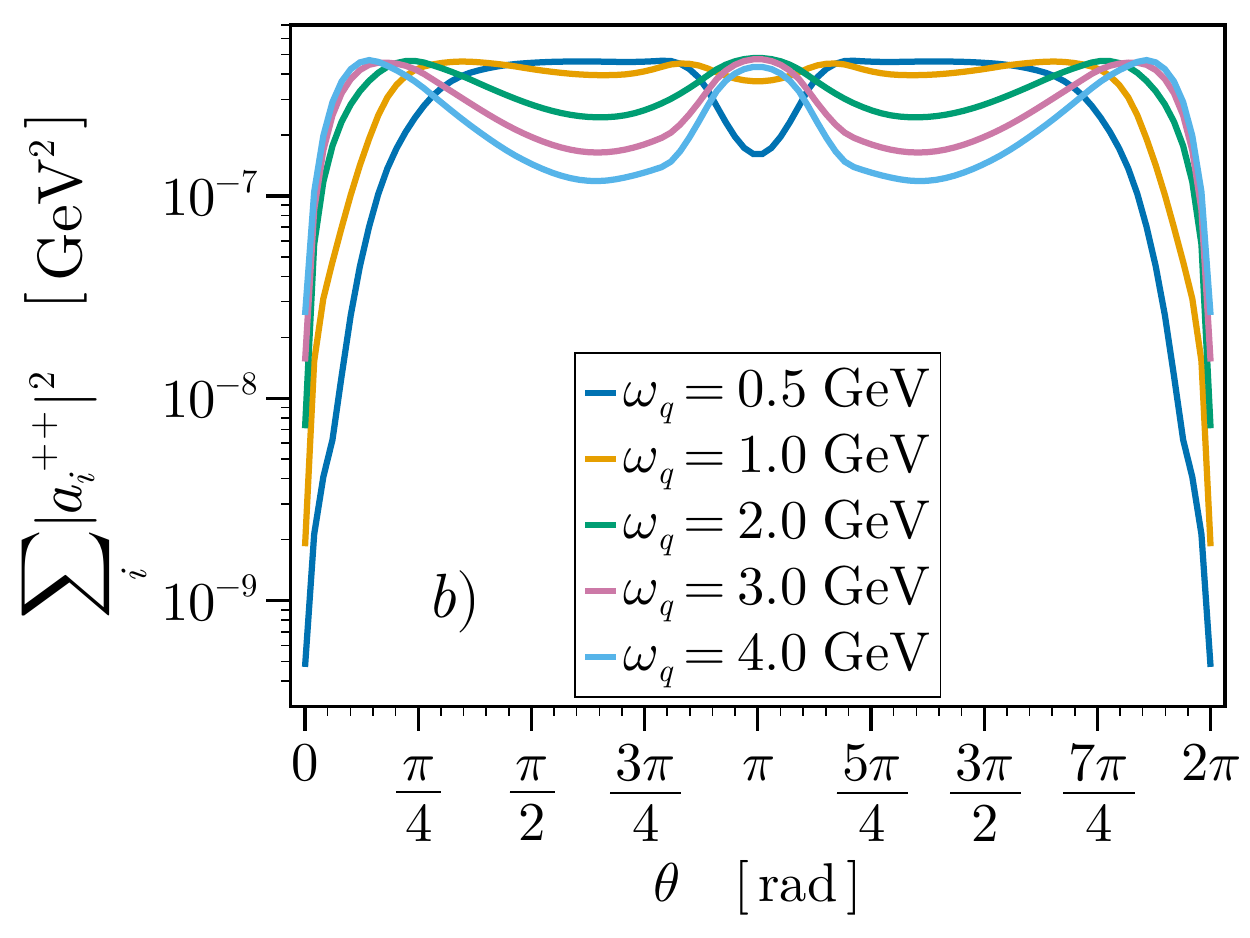} \hfill
    \includegraphics[width=0.33\textwidth]{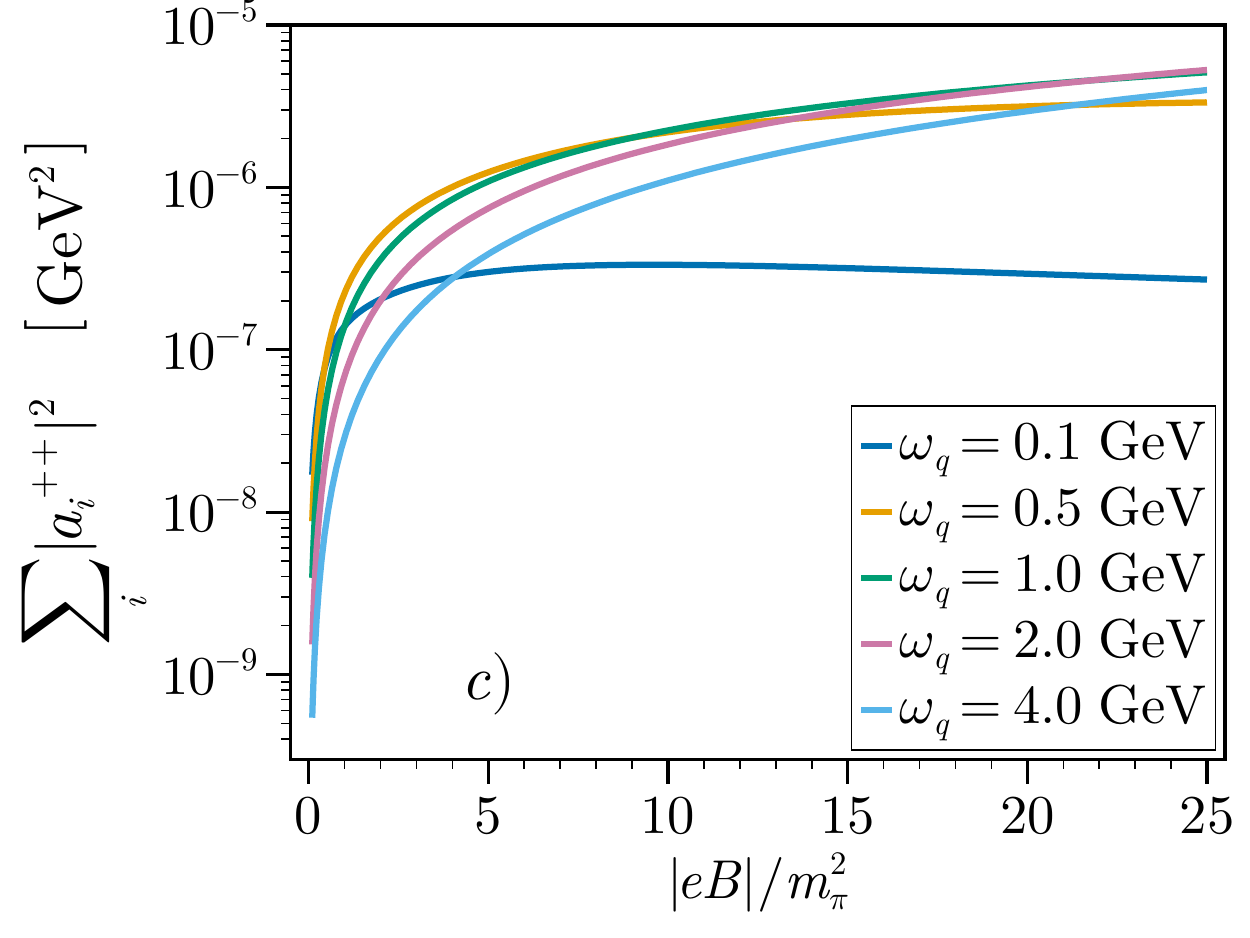}
    \caption{a) Squared matrix element for the fusion process as a function of the photon energy for different values of one of the gluons energy, for fixed values of $|eB| = 3m_\pi ^2$ and $\theta = \pi / 2$. b) Squared matrix element dependence on the angle between the photon momentum and the direction of the magnetic field, for different values of the photon energy, fixed $\omega_p = \omega_q / 2$ and $|eB| = 3m_\pi^2$. c) Squared matrix element as a function of the magnetic field strength scaled by the squared of the pion mass, for different values of the photon energy and for fixed $\omega_p = \omega_q / 2$ and $\theta = \pi / 2$. In all the plots, the matrix elements are calculated as the sum over the quark flavors $f = u,d,s$ and $\alpha_s = 0.3$.}
    \label{fig:NumVertex}
\end{figure}
\end{widetext}

The gluon splitting process can be computed using crossing symmetry
\begin{equation}
\Gamma_{gg\to\gamma} \to \Gamma_{g\to g\gamma},
\label{crossing}
\end{equation}
and carrying out the corresponding momentum changes.

\section{Photon yield and \texorpdfstring{$v_2$}{v2}}\label{sec:IV}

The invariant photon momentum distribution from gluon fusion and splitting, summed over initial and averaged over final polarizations, is obtained from the phase space integration of the squared matrix element for each process. In writing the squared matrix element, we recall that the processes are related to each other by means of the crossing symmetry, and so we can write it as
\be
{\mbox{fusion}}&=&\sum_f (2\pi)^4 \delta^{(4)}(p_1 + p_2 - q) |\Gamma_f|^2 \nonumber\\
{\mbox{splitting}}&=& \sum_f (2\pi)^4 \delta^{(4)}(p_1 - p_2 - q) |\Gamma_f|^2,
\ee
where $\Gamma_f$ is the vertex for the process $gg\to\gamma$ with the fermion $f$ as the intermediate state in the loop. Hence, the invariant momentum distribution can be written as 
\be
\omega_q \frac{\dd{N}^{\text{mag}}}{\dd[3]{q}} & \!=\! & \frac{\mathcal{VT}}{2(2\pi)^3} \int \frac{\dd[3]p_1}{(2\pi)^3 2|p_1|}\frac{\dd[3]p_2}{(2\pi)^3 2|p_2|}\, n(p_1)  \nonumber\\
& \times & \: \frac{1}{4}\sum_f \Big[n(p_2) (2\pi)^4 \delta^{(4)}(p_1 + p_2 - q)|\Gamma_f|^2\nonumber\\
& + &  \big(1+n(p_2)\big)(2\pi)^4 \delta^{(4)}(p_1 - p_2 - q)|\Gamma_f|^2 \Big],\IEEEeqnarraynumspace
\label{yield1}
\ee
where $n(p)$ represents the gluon distribution at pre-equilibrium, and $\mathcal{VT}$ is the space-time volume of the region where photons are being produced, and is approximated as $\mathcal{V} = \tfrac{4}{3}\pi R^3$, with $R = 7$ fm (corresponding to the gold nuclear radius) and $\mathcal{T} = 1.5$ fm. A complete description of photon production by gluon fusion and splitting requires knowledge of the time evolution of the gluon distribution during pre-equilibrium, as well as the space-time evolution of the parameters, such as the magnetic field and the space-time volume of the interaction region. Furthermore, it is expected that such parameters change with the collision energy and centrality. However, in this work we aim at a qualitative ballpark description of the photon spectrum and $v_2$ to show that these magnetic field induced processes do provide the overall order of magnitude and shape required by the experimental data. Hence, under these assumptions, the yield of photons due to gluon fusion and splitting is then computed as the anti-derivative of the invariant momentum distribution, which for mid-rapidity is given by
\be
\frac{1}{2\pi \omega_q} \dv{N^{\text{mag}}}{\omega_q} & = & \frac{\mathcal{VT}}{8(2\pi)^5 \omega_q} \int_0^\infty \!\!\dd{p_1} \int_0^{\pi}\!\! \dd{\theta}n(p_1) \nonumber\\
& \times & \frac{1}{4}\sum_f \Big[ n(q-p_1) |\Gamma_f|^2 _{p_2 = q - p_1} \Theta(\omega_q - \omega_p) \nonumber\\
& + & \big(1 + n(p_1 - q) \big) |\Gamma_f| ^2 _{p_2 = p_1 - q} \Theta(\omega_p - \omega_q)\Big],\nonumber\\
\IEEEeqnarraynumspace
\label{yield2}
\ee
where $\Theta(x)$ is the Heaviside function that arises from the conservation of energy and the last term explicitly shows that one of the gluons is in the final state. 

For the gluon distribution, one can consider, as in Ref.~\cite{Ayala:2017vex}, an isotropic Bose-Einstein distribution inspired by the gluon occupation expected during the evolution from the glasma toward the pre-equilibrium stage
\begin{equation}
\label{eq:distBE}
    n(\omega) = \frac{\eta}{e^{\flatfrac{\omega}{\Lambda_s}}-1},
\end{equation}
where $\eta$ represents the gluon occupation factor, and $\Lambda_s$ is the saturation momentum scale.

Another possibility for the gluon distribution has been introduced in Refs.~\cite{Kurkela:2015qoa, Ayala:2024jvc}. These works have argued that the gluon distribution at pre-equilibrium should reflect a strong anisotropy in the beam direction. This distribution can be written as~\cite{Kurkela:2015qoa, Garcia-Montero:2023lrd}
\be
\label{eq:distAnis}
    n(p_L, p_\perp) & = & \frac{2}{\lambda} \, \eta\, A \, n_0\left(\xi \tfrac{p_L}{\expval{p_T}}, \tfrac{p_\perp}{\expval{p_T}}\right), \\
    n_0(\hat{p}_L, \hat{p}_\perp) & = & \frac{1}{\sqrt{\hat{p}_\perp ^2 + \hat{p}_L ^2}} e^{-\frac{2}{3}(\hat{p}_L ^2 + \hat{p}_\perp ^2)},
\ee
where $\lambda = 4\pi N_c \alpha_s$ is the t'Hooft coupling, with $N_c = 3$ being the number of colors, $A = 5.34$, and $\eta$ is the gluon occupation factor, analogous to the one in Eq~\eqref{eq:distBE}; $p_L$ is the momentum component in the beam direction, $\xi$ is a coefficient quantifying the momentum anisotropy in the beam direction, $p_\perp$ is the momentum component in the direction transverse to the beam, and $\expval{p_T}$ is the mean transverse momentum scale. In Refs.~\cite{Kurkela:2015qoa, Garcia-Montero:2023lrd}, the anisotropy coefficient is taken to be of the order of $\xi = 10$. However, Ref.~\cite{Ayala:2024jvc} has argued that, due to the presence of the magnetic field, $\xi$ should be smaller. 

We test the sensitivity of our results to these two possible gluon distributions. To this end, we first compute the magnetic contribution to the $v_2$ coefficient. Expanding the photon azimuthal distribution in its Fourier modes, we can write
\begin{equation}
    \dv{N^\text{mag}}{\phi} = \frac{N^{\text{mag}}}{2\pi}\left[ 1 + \sum_{n=1} ^\infty 2v_n ^\text{mag} \cos(n\phi) \right],
\end{equation}
where 
\begin{equation}
    N^\text{mag} \equiv \int \dd[3]{q} \frac{\dd N^\text{mag}}{\dd[3]{q}},
\end{equation}
is the total number of photons produced by gluon fusion and splitting processes, and $\phi$ is the angle between the photon momentum and the reaction plane. Hence, the magnetic field contribution to the elliptic flow coefficient is given by
\begin{equation}
    v_2 ^\text{mag} (\omega_q) = \frac{1}{N^\text{mag}}\int_0 ^{2\pi}\dd{\phi} \cos(2\phi) \dv{N^\text{mag}}{\phi},
\end{equation}
where $\dv*{N^\text{mag}}{\phi}$ can be computed as the anti-derivative of the invariant momentum distribution. Additionally, since the matrix element depends on the angle between the photon momentum and the magnetic field, we notice that for mid-rapidity, these angles differ only by $\flatfrac{\pi}{2}$, and so the integration over $\phi$ can be replaced by an integration over $\theta$. Therefore, for mid-rapidity, the elliptic flow coefficient is given by
\be 
v_2^\text{mag} (\omega_q) & = & \frac{\mathcal{VT}}{8(2\pi)^4 N^\text{mag}} \int_0 ^{\omega_q} \!\!\!\dd{\omega_q '} \int_0 ^{2\pi} \!\!\!\dd{\theta} \int_0 ^\infty\dd{p_1} n(p_1)\nonumber \\
& \times &  \cos(2(\theta - \pi / 2))\nonumber\\
& \times & \frac{1}{4}\sum_f \Big[ n(q-p_1) |\Gamma_f|^2 _{p_2 = q-p_1} \Theta(\omega_q - \omega_p) \nonumber\\
& + &\big(1+(n(p_1 - q) \big)|\Gamma_f|^2 _{p_2 = p_1 - q} \Theta(\omega_p - \omega_q)\Big].\nonumber\\
\ee 
\begin{figure}[t]
    \centering
    \includegraphics[width=0.45\textwidth]{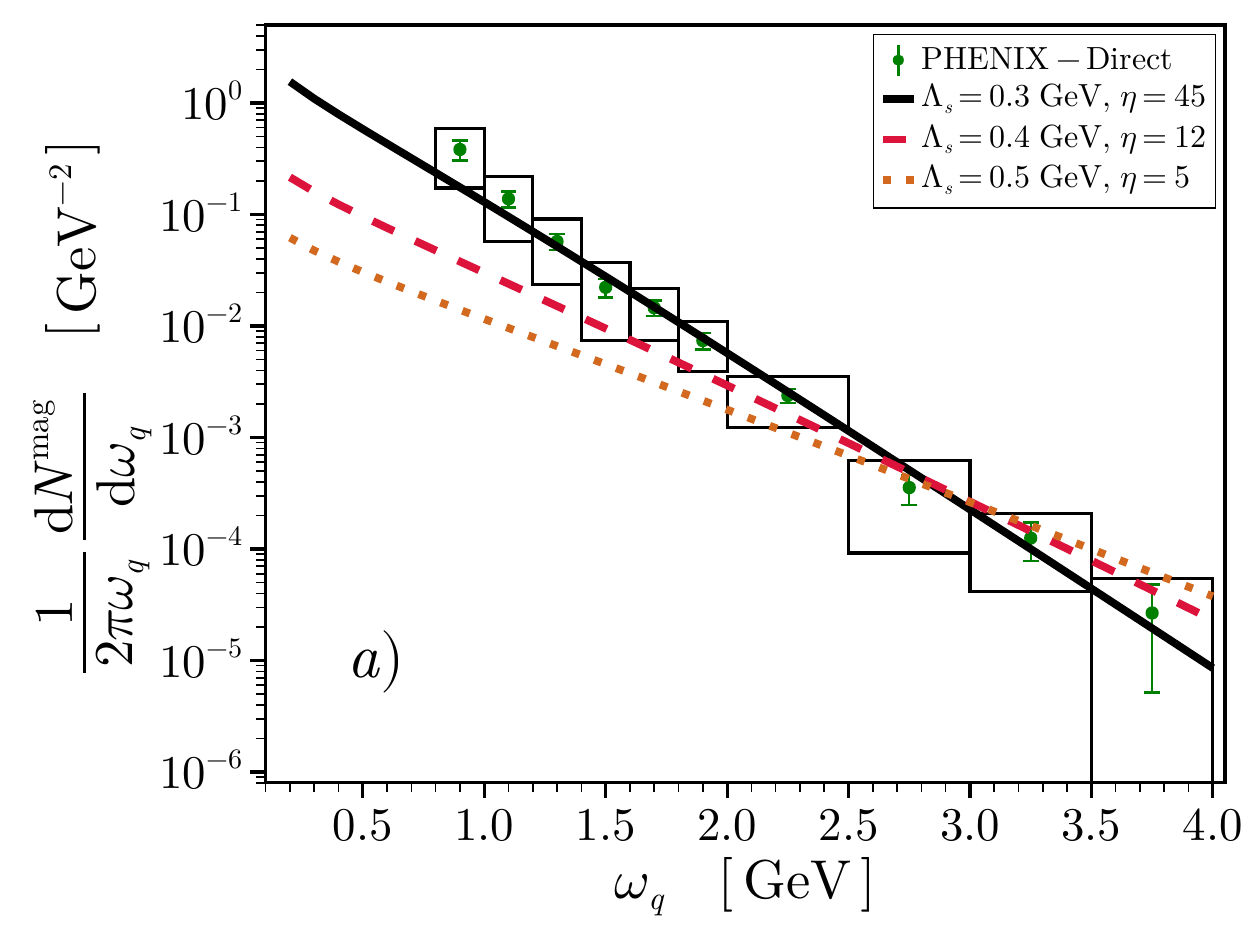} \hfill
    \includegraphics[width=0.45\textwidth]{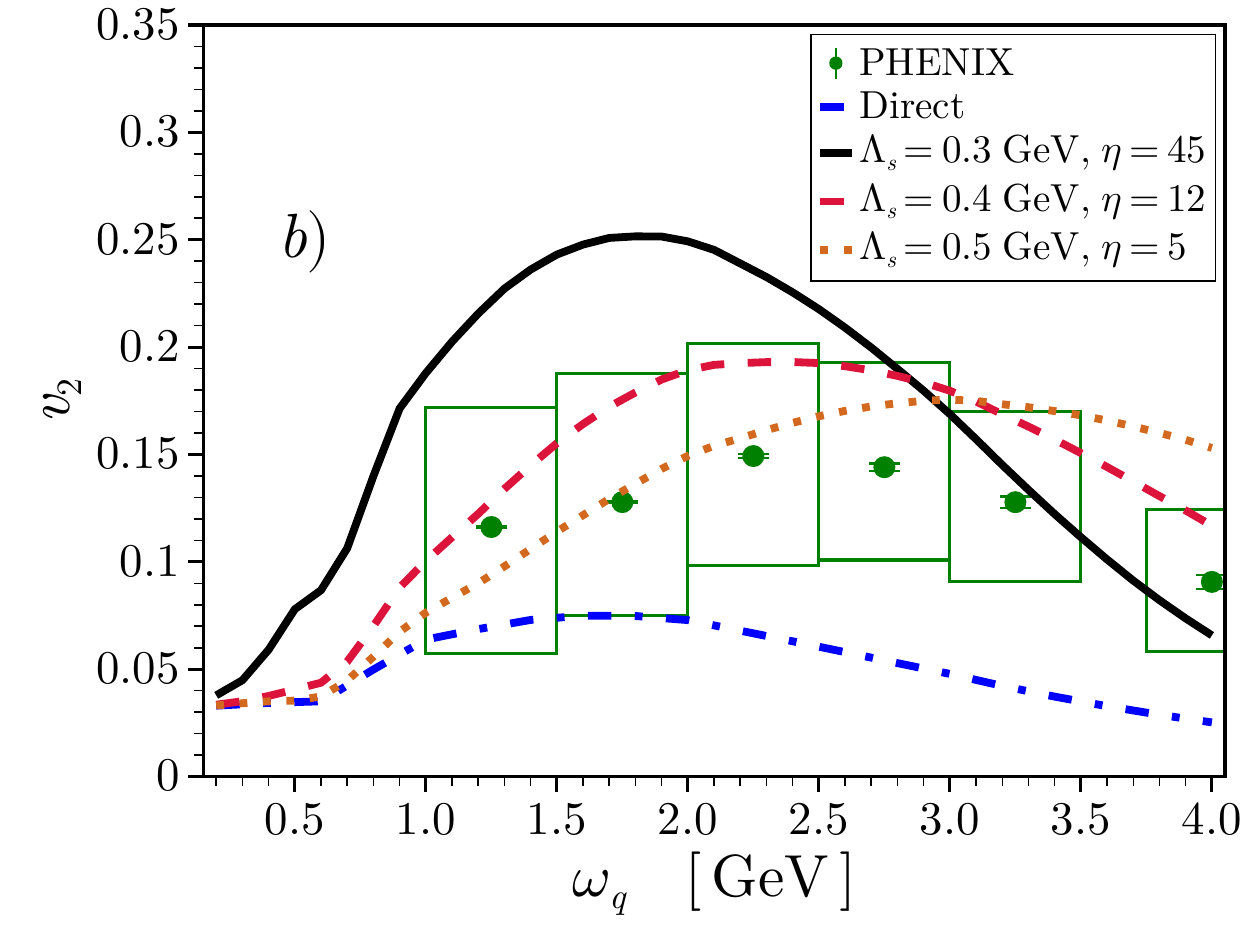}
    \caption{a) Total magnetic contribution to the photon yield (fusion plus splitting) compared to the PHENIX data of Ref.~\cite{PHENIX:2022rsx} minus the direct-photon contribution of Ref.~\cite{Gale:2021emg},  which includes prompt, thermal, and pre-equilibrium photons. b) Magnetic contribution to the elliptic flow coefficient added the contribution of direct photons of Ref.~\cite{Gale:2021emg}, compared to the PHENIX data of Ref.~\cite{PHENIX:2025ejr}. The plots are compared with data in the 20--30\% centrality class for Au+Au collisions at $\sqrt{s_{NN}} = 200$ GeV and computed for $|eB| = 3 m_\pi ^2$ using the distribution of Eq.~\eqref{eq:distBE}. In both panels, the solid black line shows the results for $\Lambda_s = 0.3$ GeV and $\eta = 45$; dashed red line shows the results for $\Lambda_s = 0.4$ GeV and $\eta = 12$; while the dotted chocolate line shows the result for $\Lambda_s = 0.5$ GeV and $\eta = 5$. In b), the direct photons' contribution is shown in the blue dash-dotted line.}
    \label{fig:ResB-E}
\end{figure}

\begin{figure}[t]
    \centering
    \includegraphics[width=0.45\textwidth]{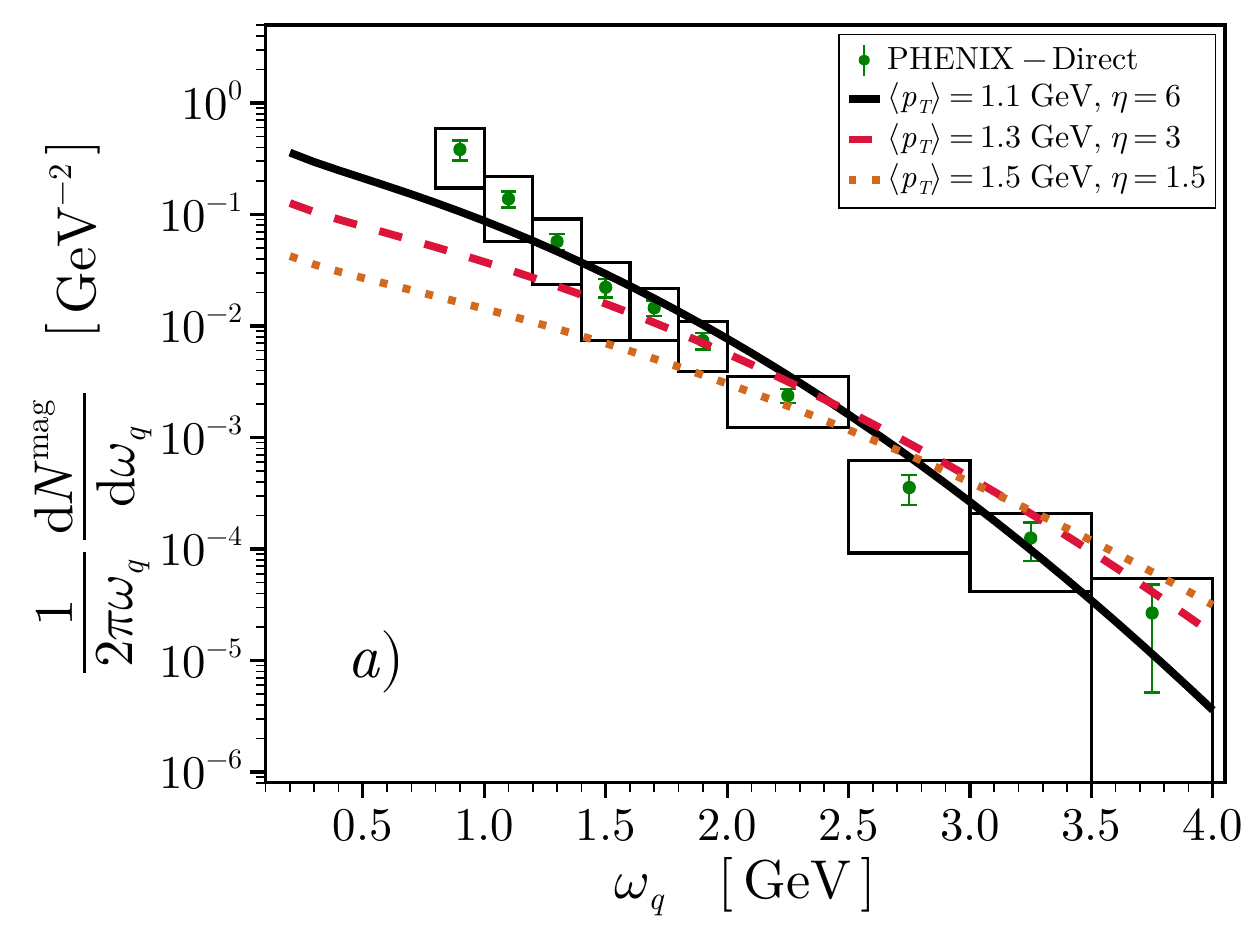} \hfill
    \includegraphics[width=0.45\textwidth]{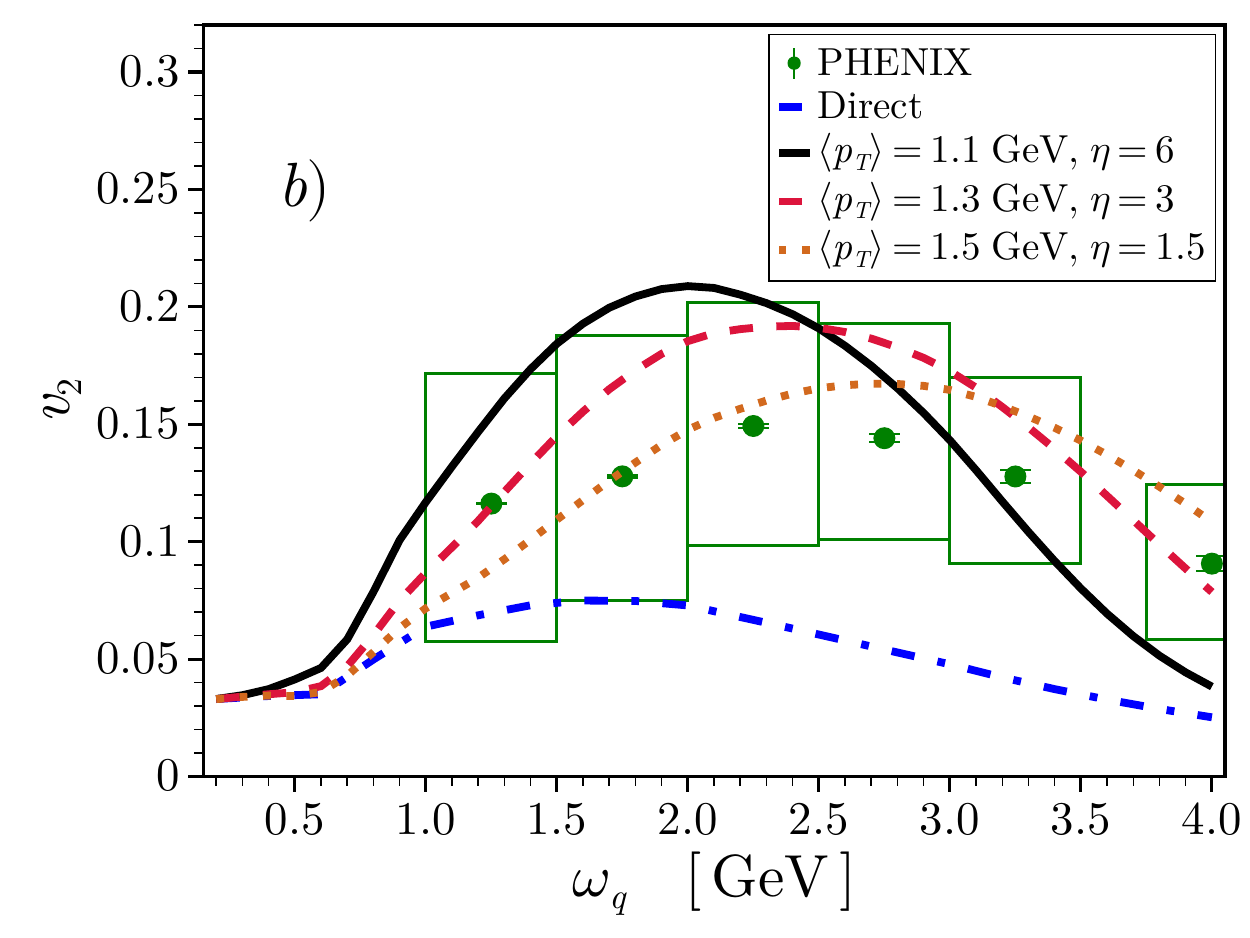}
    \caption{a) Total magnetic contribution to the photon yield (fusion plus splitting) compared to the PHENIX data of Ref.~\cite{PHENIX:2022rsx} minus the direct-photon contribution of Ref.~\cite{Gale:2021emg},  which includes prompt, thermal, and pre-equilibrium photons. b) Magnetic contribution to the elliptic flow coefficient added the contribution of direct photons of Ref.~\cite{Gale:2021emg}, compared to the PHENIX data of Ref.~\cite{PHENIX:2025ejr}. The plots are compared with data in the 20--30\% centrality class for Au+Au collisions at $\sqrt{s_{NN}} = 200$ GeV and computed for $|eB| = 3 m_\pi ^2$ using the distribution of Eq.~\eqref{eq:distAnis}. In both panels, the solid black line shows the results for $\expval{p_T} = 1.1$ GeV and $\eta = 6$; dashed red line shows the results for $\expval{p_T} = 1.3$ GeV and $\eta = 3$; while the dotted chocolate line shows the result for $\expval{p_T} = 1.6$ GeV and $\eta = 1.5$. In b), the direct photon contribution is shown in the blue dash-dotted line.}
    \label{fig:Res-Anis}
\end{figure}

With these expressions at hand, we can compare the contribution from these magnetic field driven processes with experimental measurements. We compare with data from the PHENIX collaboration for Au+Au collisions in the 20--30\% centrality class and for $\sqrt{s_{NN}} = 200$ GeV. We subtract the direct photon yield computed by the state-of-the-art  calculations of Ref.~\cite{Gale:2021emg} for the same system in the same centrality class, from the  PHENIX data~\cite{PHENIX:2022rsx}. The direct photon contribution of Ref.~\cite{Gale:2021emg} is taken as the total contribution, that takes into account prompt, thermal and pre-equilibrium photons. To numerically compute the yield coming from the magnetic field processes, we interpolate the calculated matrix element using the algorithm of Refs.~\cite{Interpolations_jl, Bezanson:2014pyv} and evaluate numerically the remaining integrals using the Suave algorithm~\cite{Hahn:2004fe, Hahn:2014fua, Bezanson:2014pyv}.

For the case of the elliptic flow coefficient, to include the contribution of the magnetic field driven processes into the direct photon contribution to $v_2$ and to compare with the data from the PHENIX collaboration~\cite{PHENIX:2025ejr}, we produced a weighted average of the magnetic contributions with the contribution from state-of-the-art calculations of Ref~\cite{Gale:2021emg} as~\cite{Ayala:2017vex}
\begin{equation}
    v_2 (\omega_q) = \frac{\dv{N^\text{mag}}{\omega_q} v_2 ^\text{mag}(\omega_q) + \dv{N^\text{direct}}{\omega_q} v_2 ^\text{direct}(\omega_q)}{\dv{N^\text{mag}}{\omega_q} + \dv{N^\text{direct}}{\omega_q}}.
\label{v2expression}
\end{equation}
We compute the magnetic contributions to the elliptic flow coefficient by following a numerical procedure analogous to that used to calculate the yield.

\begin{figure}[t]
    \centering
    \includegraphics[width=0.45\textwidth]{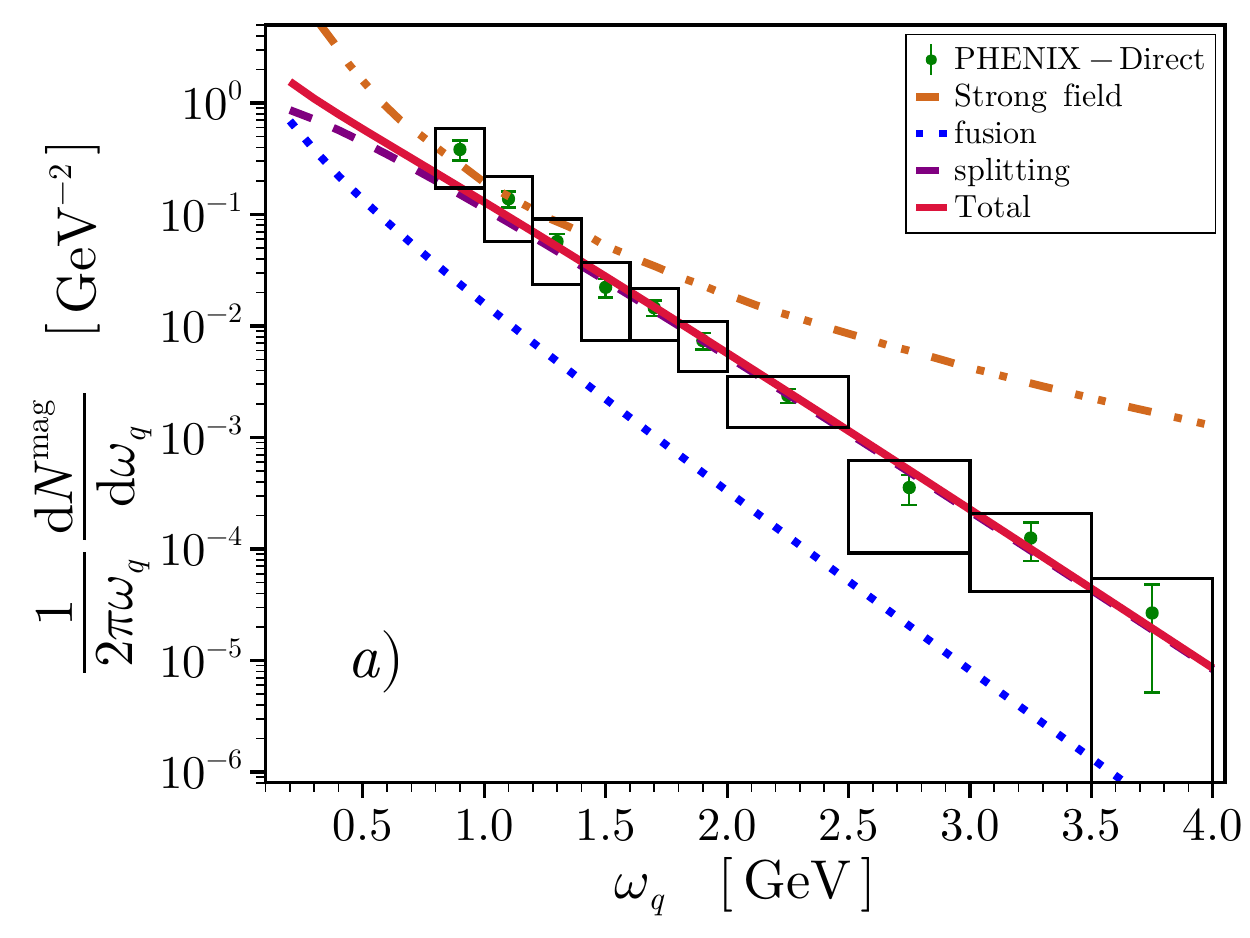} \hfill
    \includegraphics[width=0.45\textwidth]{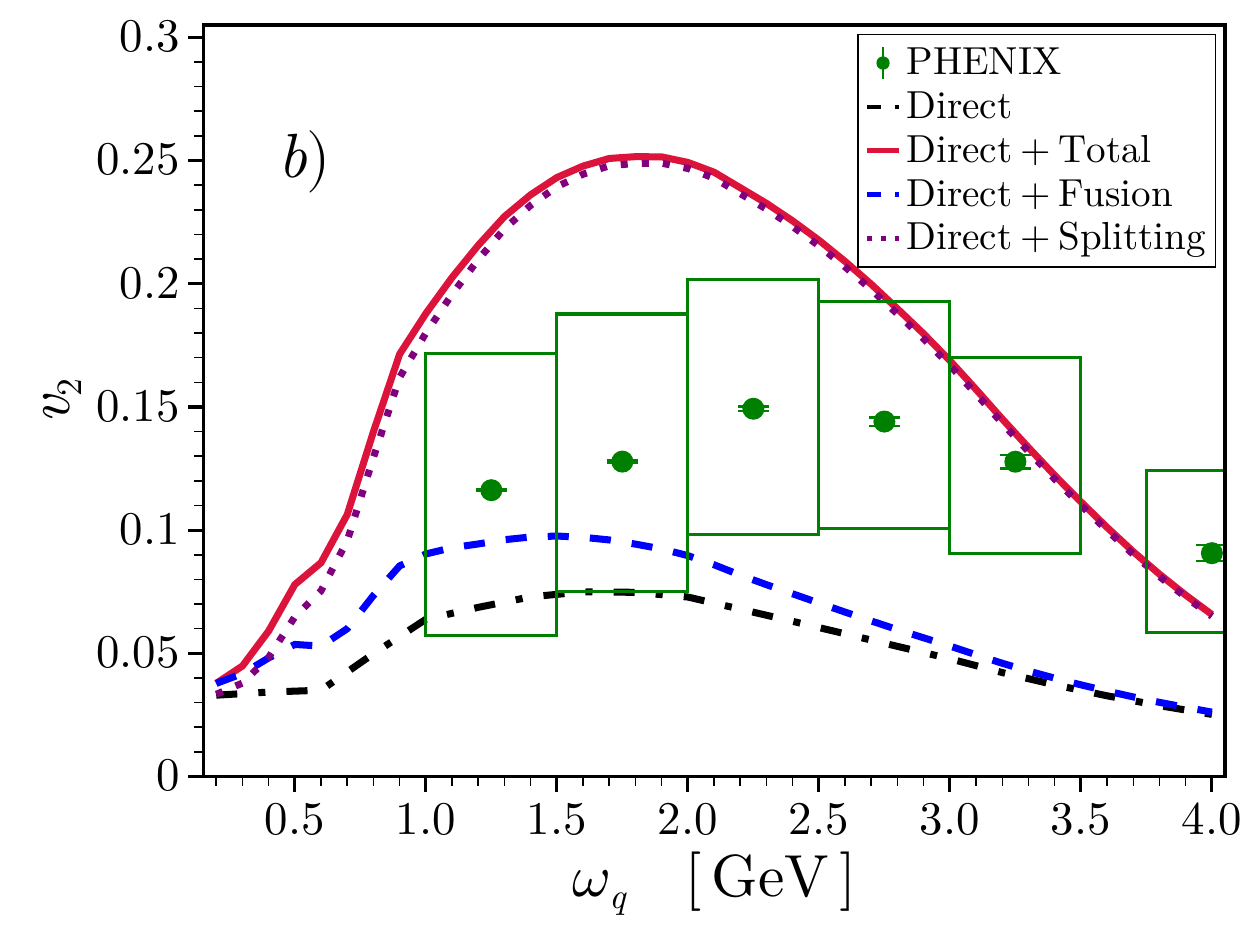}
    \caption{a) Photon yield obtained from the gluon fusion (blue dotted line), splitting (purple dashed line) and their sum (red solid line) for $|eB| = 3\,m_\pi^2$ using the distribution of Eq.~\eqref{eq:distBE} with $\Lambda_s = 0.3$ GeV and $\eta = 45$, compared to the PHENIX data of Ref.~\cite{PHENIX:2022rsx} minus the direct-photon contribution of Ref.~\cite{Gale:2021emg},  which includes prompt, thermal, and pre-equilibrium photons. The strong magnetic field result of Ref.~\cite{Ayala:2017vex} (chocolate dash-dot-dotted line) is also shown. b) Weighted average $v_2$ for the same set of parameters with the contribution of direct photons of Ref.~\cite{Gale:2021emg}, compared with the PHENIX data of Ref.~\cite{PHENIX:2025ejr}. The red solid line represents the total contribution, while the dash-dotted black line represents the direct contribution. Also shown are the weighted average of the fusion (blue dashed line) and of splitting (purple dotted line) contributions. The plots are compared with data in the 20--30\% centrality class for Au+Au collisions at $\sqrt{s_{NN}} = 200$ GeV.}
    \label{fig:Best-BE}
\end{figure}

Figure~\ref{fig:ResB-E} shows the total contribution of the magnetic field driven processes to the photon distribution and elliptic flow coefficient for a magnetic field strength $|eB| = 3m_\pi ^2$ and different values of $\Lambda_s$ and $\eta$, using the distribution of Eq.~\eqref{eq:distBE}. The values of $\Lambda_s$ and $\eta$ were independently chosen in order to give a good description of the PHENIX data. Overall, there is good agreement between the obtained yield and the data. However, in the case of the $v_2$, the smallest value of $\Lambda_s$ overestimates the data. Notice that in all these cases, the required values of $\Lambda_s$ are small. We interpret this result as a signal that, for a Bose-Einstein-type initial gluon distribution, the occupation number is dominated by very soft gluons.  Recall that in the context of the color glass condensate, the gluon distribution function for momenta similar to the gluon saturation scale must behave parametrically as
\begin{equation}
    n(p\sim Q_s) \sim \frac{1}{\alpha_s}.
\end{equation}
Therefore, for typical values of $\alpha_s \sim 0.1 - 0.3$, $\eta$ must be restricted to the order of $\mathcal{O}(10)$. Hence, all the considered values of $\eta$ in  Fig.~\ref{fig:ResB-E} are within this expected order of magnitude.
\begin{figure}
    \centering
    \includegraphics[width=0.45\textwidth]{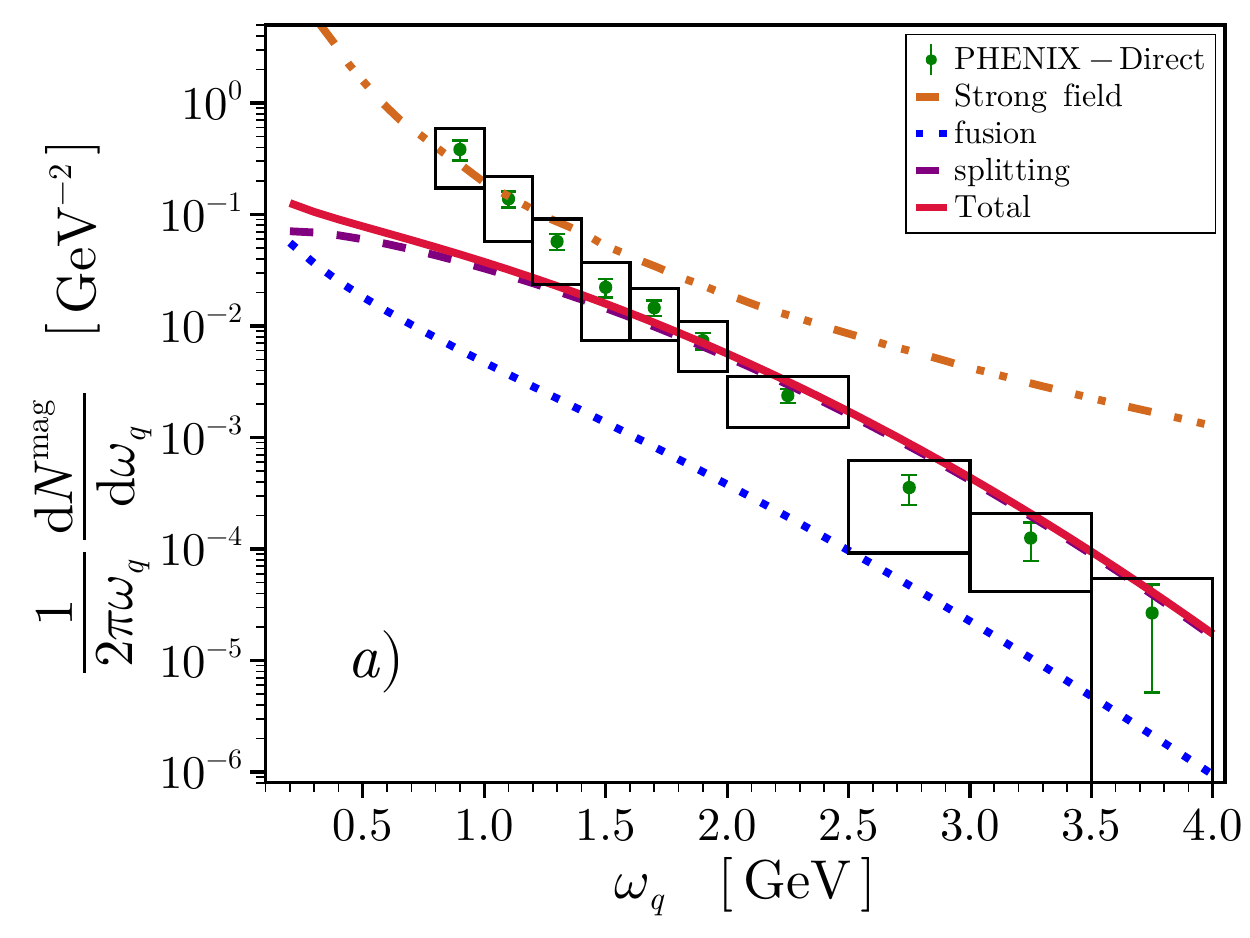} \hfill
    \includegraphics[width=0.45\textwidth]{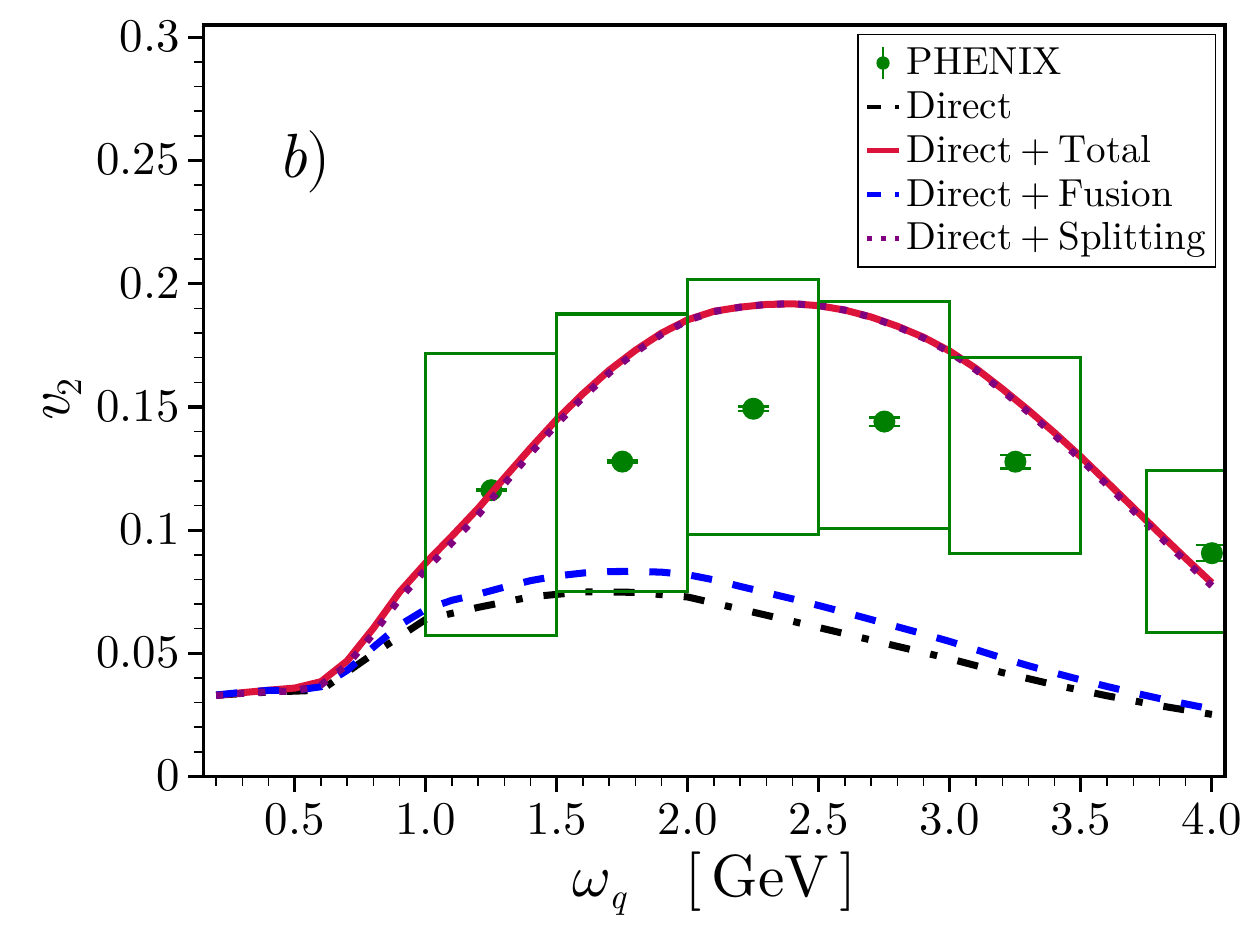}
    \caption{a) Photon yield obtained from gluon fusion (blue dotted line), splitting (purple dashed line) and their sum (red solid line) for $|eB| = 3\,m_\pi^2$ using the distribution of Eq.~\eqref{eq:distAnis} with $\expval{p_T} = 1.3$ GeV and $\eta = 3$, compared to the PHENIX data of Ref.~\cite{PHENIX:2022rsx} minus the direct-photon contribution of Ref.~\cite{Gale:2021emg},  which includes prompt, thermal, and pre-equilibrium photons. The strong magnetic field result of Ref.~\cite{Ayala:2017vex} (chocolate dash-dot-dotted line) is also shown. b) Weighted average $v_2$ for the same set of parameters with the contribution of direct photons of Ref.~\cite{Gale:2021emg}, compared with the PHENIX data of Ref.~\cite{PHENIX:2025ejr}. The red solid line represents the total contribution, while the dash-dotted black line represents the direct contribution. Also shown are the weighted average of the fusion (blue dashed line) and of splitting (purple dotted line) contributions. The plots are compared with data in the 20--30\% centrality class for Au+Au collisions at $\sqrt{s_{NN}} = 200$ GeV.}
    \label{fig:Best-Anis}
\end{figure}

Figure~\ref{fig:Res-Anis} shows the total contribution of the magnetic field driven processes to the photon distribution and elliptic flow coefficient using the distribution of Eq.~\eqref{eq:distAnis} for $|eB|=3m_\pi^2$ with an anisotropy coefficient $\xi = 1$, for different values of $\expval{p_T}$, which are similar to the gluon saturation scale $Q_s$ and $\eta$, respectively. In all the described cases, there is good agreement with the PHENIX data, except perhaps for the lowest energy bins, with $\eta$ taking values within the expected order of magnitude.
\begin{figure}[t!]
    \centering
    \includegraphics[width=0.45\textwidth]{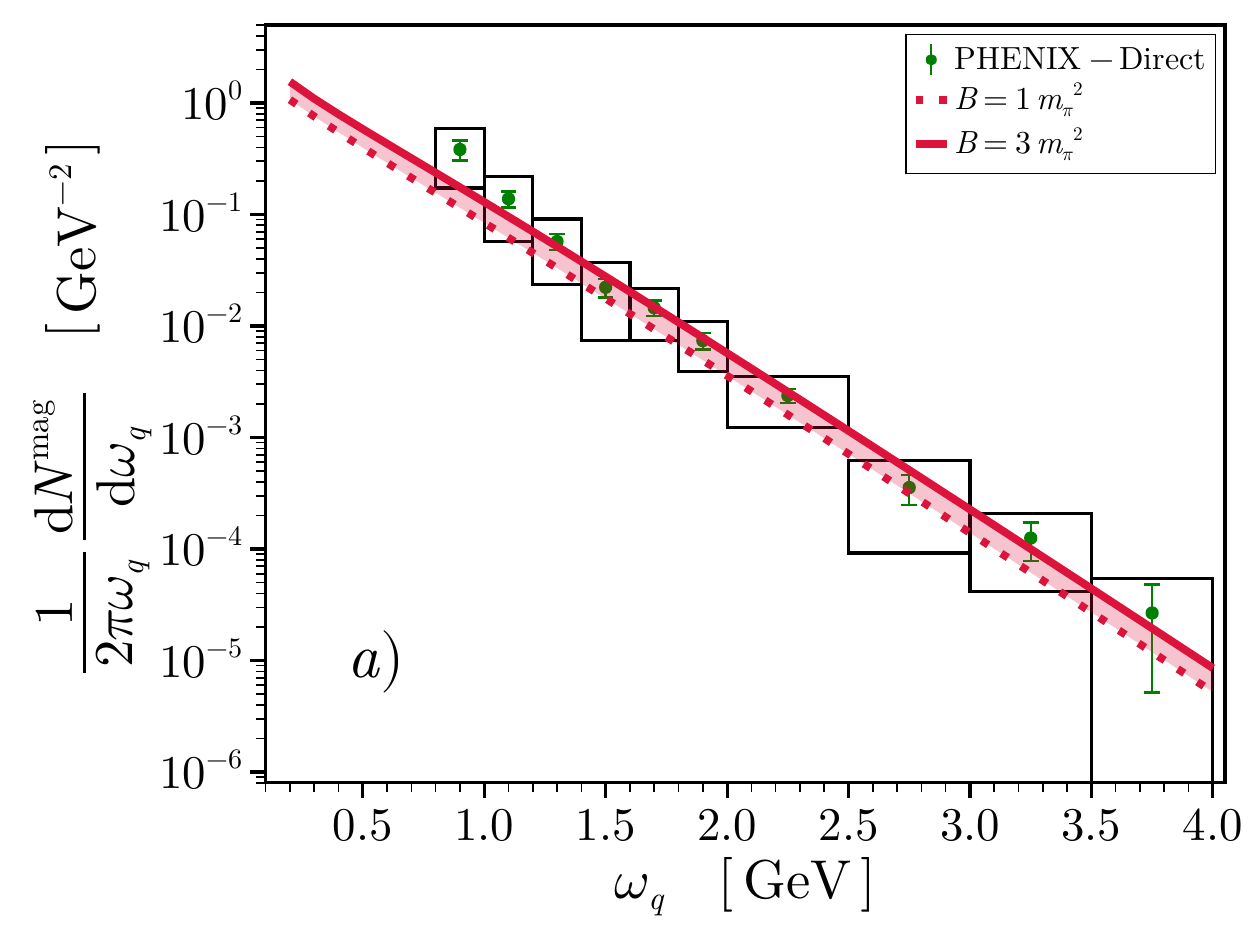} \hfill
    \includegraphics[width=0.45\textwidth]{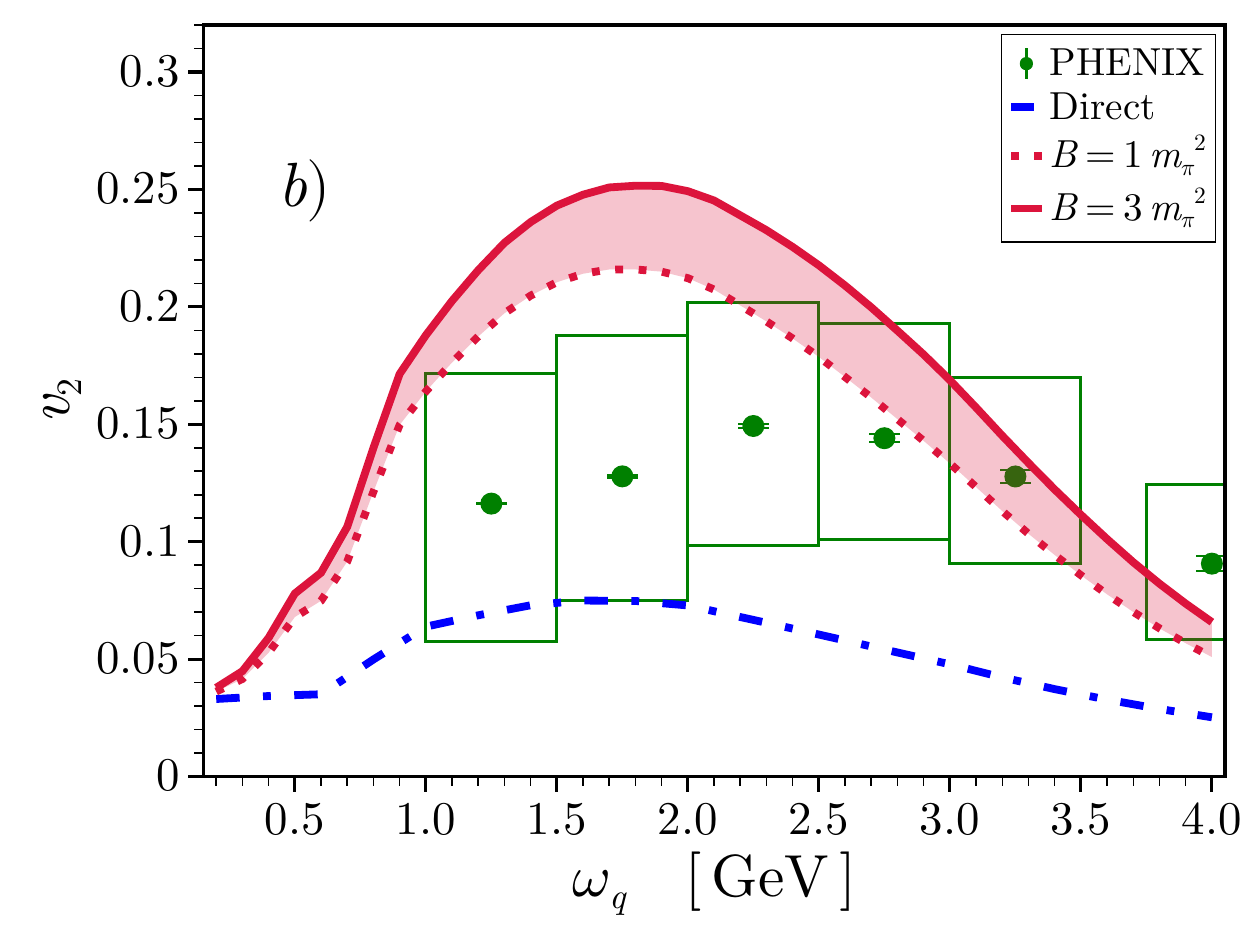}
    \caption{a) Total magnetic photon yield compared to the PHENIX data from Ref.~\cite{PHENIX:2022rsx} minus the direct-photon contribution of Ref.~\cite{Gale:2021emg},  which includes prompt, thermal, and pre-equilibrium photons. b) Weighted average elliptic flow coefficient, which includes the contribution of direct photons from Ref.~\cite{Gale:2021emg}, compared to the PHENIX data of Ref.~\cite{PHENIX:2025ejr}; using the distribution of Eq.~\eqref{eq:distBE}. In both panels, $\Lambda_s = 0.3$ GeV and $\eta = 45$ were used; the dotted red line shows the result for $|eB| = m_\pi^2$ and the solid red line shows the result for $|eB| = 3 m_\pi^2$. In b), the direct photons contribution is shown with the blue dash-dotted line. The plots are compared with data in the 20--30\% centrality class for Au+Au collisions at $\sqrt{s_{NN}} = 200$ GeV.}
    \label{fig:B-E_diff-B}
\end{figure}

Figures~\ref{fig:Best-BE} and~\ref{fig:Best-Anis} show the independent contributions to the photon distribution from gluon fusion and splitting, together with their sum. Also shown are the elliptic flow coefficients. For these figures, we use the gluon distributions of Eqs.~\eqref{eq:distBE} and~\eqref{eq:distAnis}, respectively, and a magnetic field strength $|eB|=3m_\pi^2$, with the rest of the parameters as indicated in the figures. It can be seen that for both distributions, the contribution of the splitting process dominates for the yield and the $v_2$, except for extremely low energies. Recall that, although the fusion and splitting rates are related by crossing symmetry, Eq.~(\ref{crossing}), the corresponding contributions to the photon production rate involve different kinematic domains and, importantly, different statistical factors. As shown explicitly in Eqs.~(\ref{yield1}) and~(\ref{yield2}), fusion is weighted by $n(p_1)n(p_2)$, since two gluons must be present in the initial state, whereas splitting is weighted by $n(p_1)[1+n(p_2)]$, since only one initial gluon is required, and the final-state gluon gives rise to a Bose-enhancement factor. Consequently, as the gluon occupation number decreases with increasing momentum, the fusion contribution is suppressed approximately as $n^2$, while the splitting contribution is suppressed only as $n$ when $n\ll 1$. This provides the main physical reason for the dominance of splitting over most of the photon-energy range shown in Figs.~\ref{fig:Best-BE} and~\ref{fig:Best-Anis}. At very low energies, where the gluon occupation is large and $1+n\simeq n$, this statistical advantage becomes less important, and the detailed phase space and momentum dependence of the crossed matrix element determines the relative magnitude of the two contributions.

The obtained yield and elliptic flow coefficient in both cases give a better description of PHENIX data than the findings of Ref.~\cite{Ayala:2017vex}, where a strong magnetic field approximation was used. On general grounds, for the photon yield, the splitting and fusion magnetic field driven processes become more important at intermediate photon energies, and thus a better agreement with the data is expected for energies between $1.5 < \omega_q < 4$ GeV. Both gluon distributions, Eqs.~(\ref{eq:distBE}) and~(\ref{eq:distAnis}) can provide a simultaneous qualitative description of the photon yield and elliptic flow within the explored parameter ranges. The relative agreement with the data depends on the observable and on the chosen parameters; therefore, without a dedicated statistical analysis we refrain from assigning a quantitative preference to either distribution.
\begin{figure}
    \centering
    \includegraphics[width=0.45\textwidth]{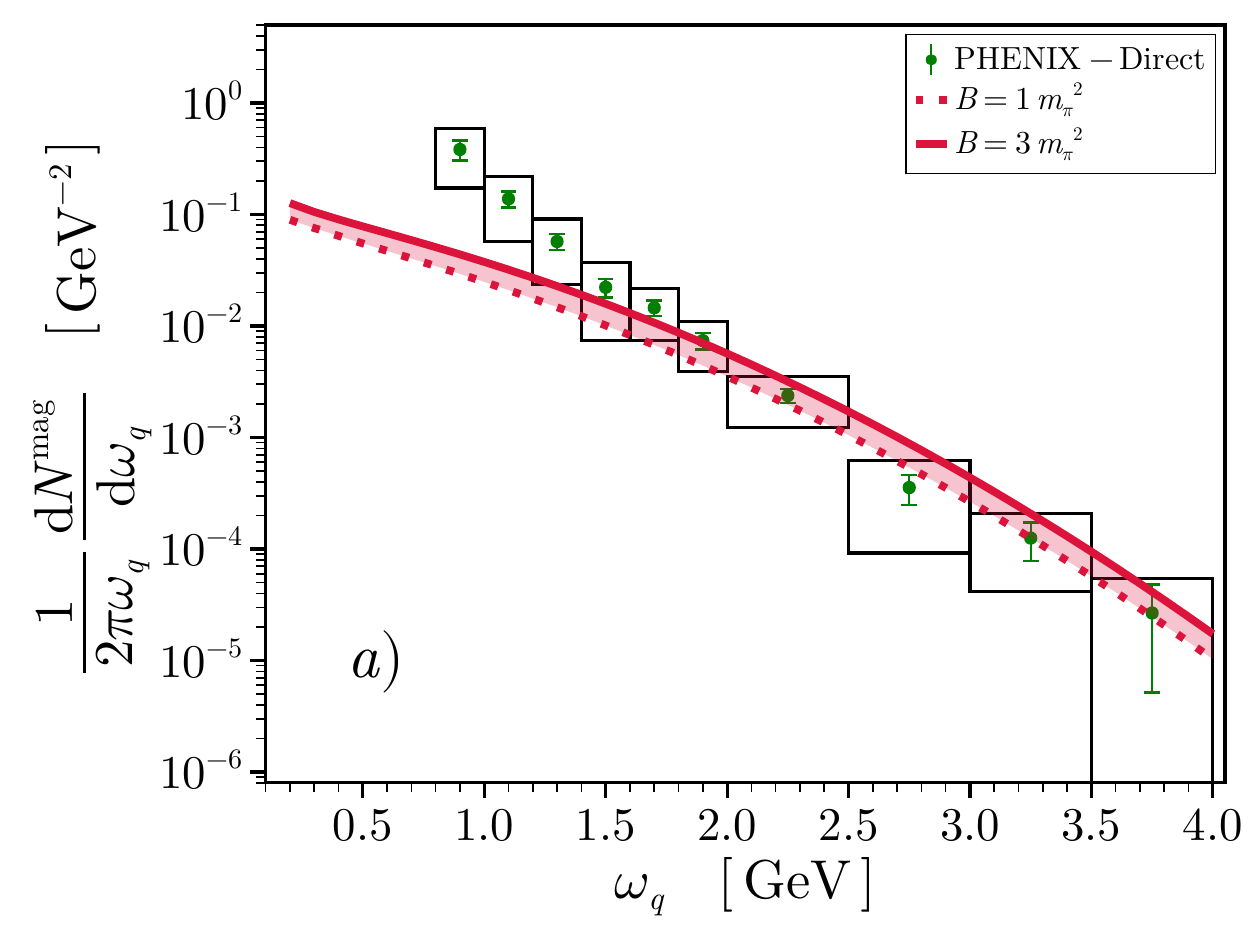} \hfill
    \includegraphics[width=0.45\textwidth]{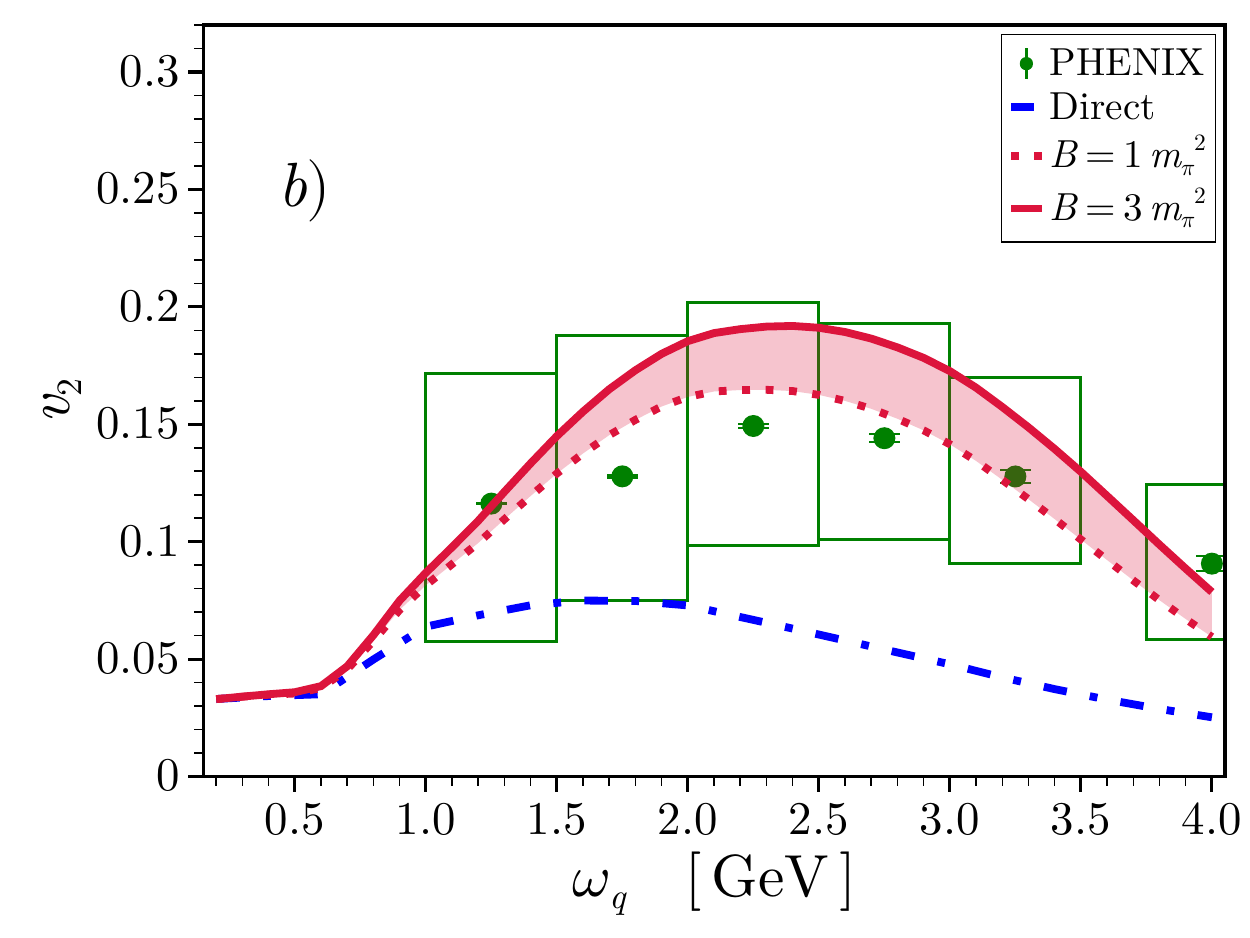}
    \caption{a) Total magnetic photon yield compared to the PHENIX data of Ref.~\cite{PHENIX:2022rsx} minus the direct-photon contribution of Ref.~\cite{Gale:2021emg},  which includes prompt, thermal, and pre-equilibrium photons. b) Weighted average elliptic flow coefficient, including the contribution of direct photons from Ref.~\cite{Gale:2021emg}, compared to the PHENIX data of Ref.~\cite{PHENIX:2025ejr}; using the distribution of Eq.~\eqref{eq:distAnis}. In both plots, $\xi = 1$, $\expval{p_T} = 1.1$ GeV and $\eta = 3$ were used; the dotted red line shows the result for $|eB| = m_\pi^2$ and the solid red line shows the result for $|eB| = 3 m_\pi^2$. In b), the direct photon contribution is shown by the blue dash-dotted line. The plots are compared with data in the 20--30\% centrality class for Au+Au collisions at $\sqrt{s_{NN}} = 200$ GeV.}
    \label{fig:Anis_diff-B}
\end{figure}

\begin{figure}[t!!!]
    \centering
    \includegraphics[width=0.45\textwidth]{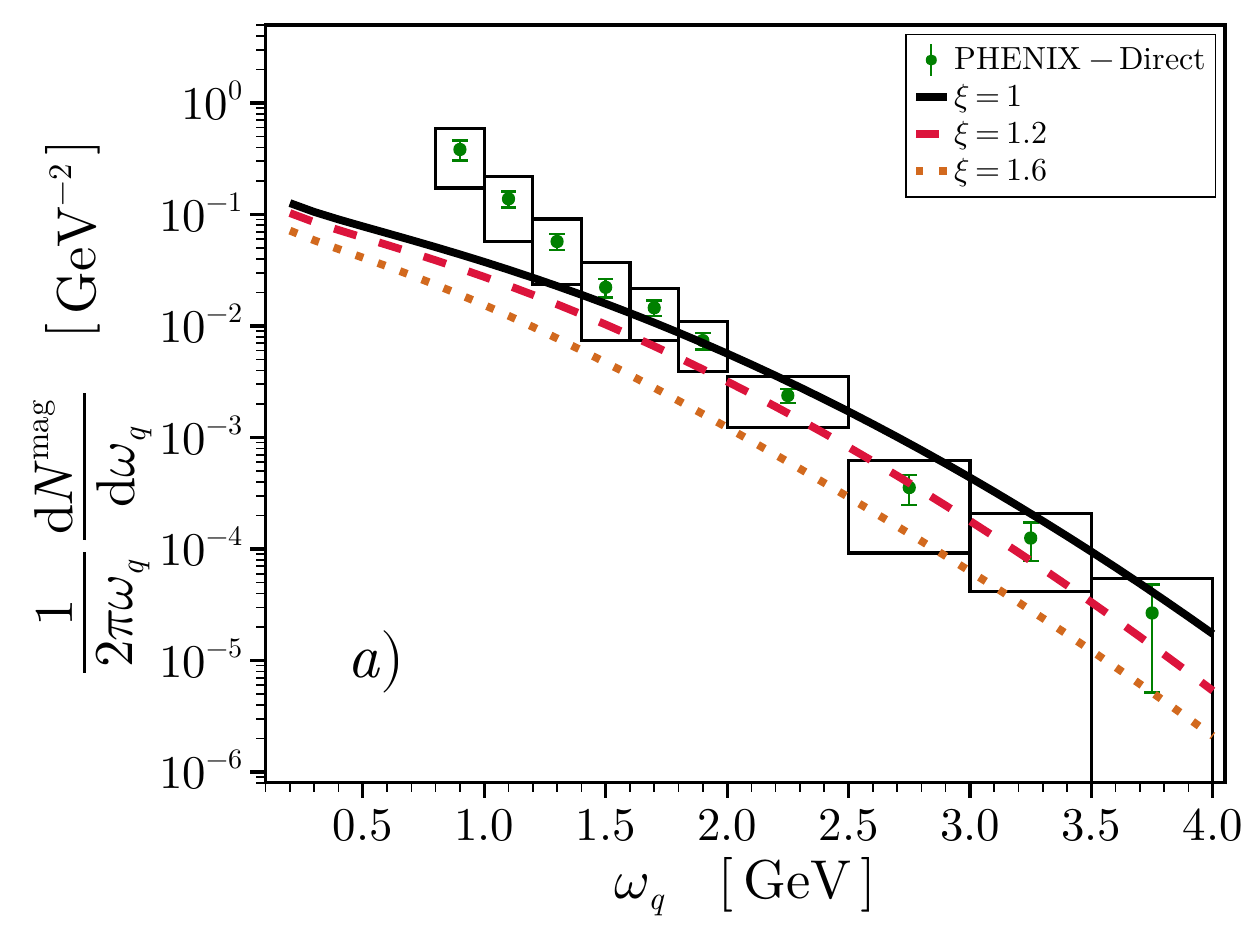} \hfill
    \includegraphics[width=0.45\textwidth]{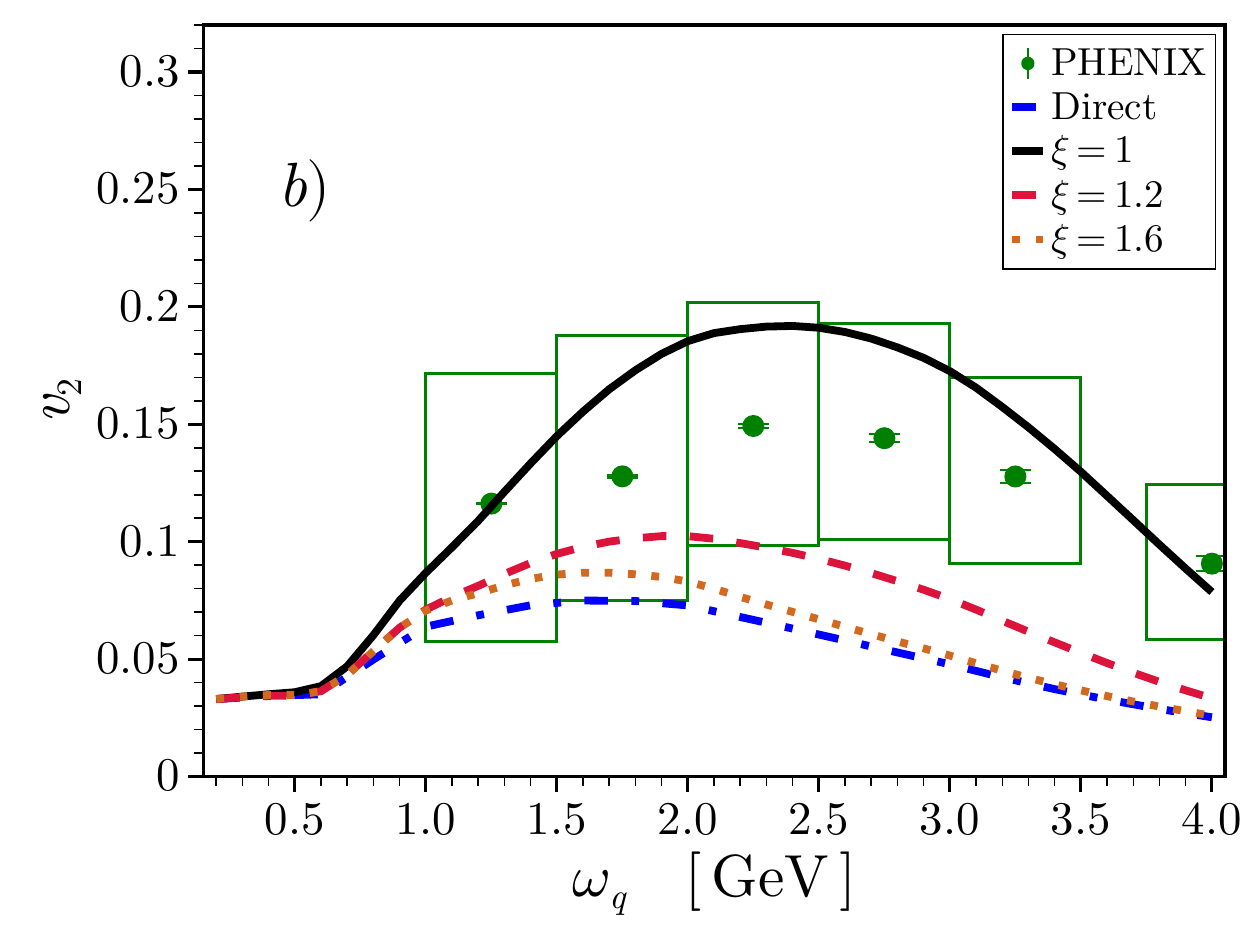}
    \caption{a) Total photon yield compared to the PHENIX data of Ref.~\cite{PHENIX:2022rsx} minus the direct-photon contribution of Ref.~\cite{Gale:2021emg},  which includes prompt, thermal, and pre-equilibrium photons. b) Weighted average elliptic flow coefficient, including the contribution of direct photons from Ref.~\cite{Gale:2021emg}, compared to the PHENIX data of Ref.~\cite{PHENIX:2025ejr}; for the case of $|eB| = 3 m_\pi ^2$, $\expval{p_T} = 1.1$ GeV and $\eta = 3$ using the distribution of Eq.~\eqref{eq:distAnis}. The solid black line shows the result for $\xi = 1$, the dashed red line shows the result for $\xi = 1.2$ and the dotted chocolate line shows the result for $\xi = 1.6$. In b), the direct photons contribution is shown as the blue dash-dotted line. The plots are compared with data in the 20--30\% centrality class for Au+Au collisions at $\sqrt{s_{NN}} = 200$ GeV.}
    \label{fig:Anis_diff-xi}
\end{figure}

\begin{table}[]
    \centering
    \begin{tabular}{c|c|c}
    \noalign{\hrule height 1.1pt}
        Parameter & Range / Value & Physical meaning \\
        \noalign{\hrule height 1.1pt}
        $\alpha_s$ & 0.3 & strong coupling \\ \hline
        $|eB|$ & $1-3\; m_\pi ^2$ & magnetic field strength \\ \hline
        $\mathcal{VT}$ & 2155 fm${}^4$ & space-time volume of interaction region\\ \hline
        $\Lambda_s$ & 0.3 -- 0.5 GeV & saturation scale \\ \hline
        $\expval{p_T}$ & 1.1 -- 1.5 GeV & mean gluon transverse momentum \\ \hline
        $\eta$ & 1.5 -- 45 & gluon occupancy factor \\ \hline
        $\xi$ & 1 -- 1.6 & gluon momentum anisotropy coefficient \\
        \noalign{\hrule height 1.1pt}
    \end{tabular}
    \caption{Summary of numerical values of the parameter used throughout this work. Each parameter is presented with the value or range of values that it takes, together with its physical interpretation.}
    \label{tab:ParamSumm}
\end{table}

Figures~\ref{fig:B-E_diff-B} and~\ref{fig:Anis_diff-B} explore the effect of the magnetic field on the photon yield and $v_2$, for the gluon distribution of Eq.~\eqref{eq:distBE} and~\eqref{eq:distAnis}, respectively. The effect of the magnetic field is to increase the yield and the $v_2$ for both distributions. However, as a function of the photon energy, the effect is non-linear for the case of the $v_2$, whereas for the yield, it is merely a displacement. Notice that, as already shown in Fig.~\ref{fig:NumVertex}(c), the squared matrix element does not grow indefinitely with the magnetic-field strength: it increases rapidly and subsequently approaches a plateau, with the onset of this behavior depending on the photon energy. Consequently, over the interval $|eB|=(1-3) m_\pi^2$, portions of the relevant phase space already exhibit a relatively weak sensitivity to further increases of the field. Moreover, $v_2$ is a normalized angular moment of the photon distribution and, in the comparison with data, it is combined with the direct-photon contribution through the yield-weighted expression of Eq.~(\ref{v2expression}). Its dependence on $B$ is therefore not expected to follow a simple power law.

On the other hand, Fig.~\ref{fig:Anis_diff-xi} shows the effect of an anisotropy in the longitudinal momentum. As can be seen, the increase in the momentum anisotropy in the beam direction decreases the magnetic contribution to the photon yield and the elliptic flow coefficient. For values of the anisotropy coefficient $\xi \gtrsim 1.7$, the contribution from the studied magnetic processes becomes negative for certain values of photon energy. This happens since, for a large initial expansion along the beam axis, emission is significant only for photons with small values of $p_L$. Therefore, the anisotropy parameter cannot be arbitrarily large in order to describe the observed photon elliptic flow coefficient. Notice that a better agreement between the PHENIX data and the magnetic field contributions is found in the isotropic ($\xi = 1$) case.  Table~\ref{tab:ParamSumm} summarizes the parameters used throughout this work, together with either the values or the range of values that were used for each parameter and its physical meaning.

\section{Summary and conclusions}\label{sec:V}

In this work, we have computed the two-gluon one-photon vertex at one-loop order in the presence of a constant magnetic field, without resorting to any further approximations. This effective vertex is written in terms of the tensor basis developed in Ref.~\cite{Ayala:2024ucr}, which takes into account all its symmetry properties and, for on-shell bosons, reduces to just three independent tensor structures. The obtained expressions can be evaluated numerically. From the obtained matrix element, we computed its contribution to the direct photon yield, as well as the elliptic flow coefficient, through the gluon fusion and splitting processes, with the latter being the dominant contribution within the explored energy range. The obtained results were compared to the data reported by the PHENIX collaboration for Au+Au collisions at $\sqrt{s_{NN}} = 200$ GeV in the 20--30\% centrality class. Since our aim was to obtain a ballpark description of the magnitude and shape of the photon yield and elliptic flow coefficient, no further attempt to compare with data from other centrality classes was made since this would require a more detailed modeling of the centrality dependent space-time volume and possibly of the evolution of the gluon phase space occupation factor and magnetic field strength with time.  For the chosen set, we found good agreement between the gluon fusion and splitting processes, together with the state-of-the-art calculations and the experimental data, particularly when we make use of the gluon distribution of Refs.~\cite{Kurkela:2015qoa, Garcia-Montero:2023lrd}, without anisotropy in the longitudinal direction of the gluon momentum.  Nonetheless, it is possible to obtain a positive contribution to the photon elliptic flow coefficient with a very small momentum anisotropy. Overall, the obtained shape for the $v_2$ shows the data trend and order of magnitude, features that are difficult to obtain by other approaches. For the representative parameter sets explored here, the magnetic contribution provides a good qualitative description over a significant part of the intermediate photon-energy region. The detailed level of agreement depends on the parameters characterizing the pre-equilibrium gluon distribution. The results are encouraging and motivate exploring the consequences of including a more detailed space-time evolution of the magnetic field as well as other processes possibly present during pre-equilibrium. These ingredients are currently being explored, and the results will soon be reported elsewhere.

\section*{Acknowledgments}

The authors thank J.-F. Paquet and J.D. Castaño-Yepes for kindly sharing their data and for very useful conversations, M.E. Tejeda-Yeomans for her useful suggestions for the numerical calculation of the matrix element, and to L. Yan, G. David, N. Scoccola, and R.L.S. Farias for their comments on the manuscript. S.B.-L. and J.J.M.-S. each acknowledge the financial support of fellowships granted by Secretar\'\i a de Ciencia, Humanidades, Tecnolog\'\i a e Innovaci\'on (SECIHTI) M\'exico  as part of the Sistema Nacional de Posgrados. A.A. thanks the colleagues and staff of Universidade de São Paulo, of Instituto de Física Teórica, UNESP and of Universidade Cidade de São Paulo for their kind hospitality during a sabbatical stay. A.A. also acknowledges support from the PASPA program of the Dirección General de Asuntos del Personal Académico (DGAPA) of the Universidad Nacional Autónoma de México (UNAM) for the sabbatical stay during which part of this research was carried out. Support for this work has been received in part by DGAPA-PAPIIT-UNAM grant number AG100826, and from Secretar\'\i a de Ciencia, Humanidades, Tecnolog\'\i a e Innovaci\'on (SECIHTI) M\'exico grant numbers CIORGANISMOS-2025-17, CBF-2025-G-1718, and CBF-2026-465. This study was financed in part by the São Paulo Research Foundation (FAPESP), Brazil. Process Numbers 2023/08826-7 and 2024/18493-8

\appendix{}
\begin{widetext}
\section{Explicit expressions for the traces in the vertex}\label{Appendix:traces}
The traces over Dirac space corresponding to diagrams $\mathcal{A}$ and $\mathcal{B}$ of Fig.~\ref{fig:Fusion} are defined as
\be
\mathcal{T}_{\mathcal{A} 1} ^{\mu\nu\alpha} + \mathcal{T}_{\mathcal{B} 1} ^{\mu\nu\alpha} & = & \Tr[\gamma^\mu \slashed{\mathcal{A}}_a \gamma^\alpha \slashed{\mathcal{A}}_b \gamma^\nu \slashed{\mathcal{A}}_c] + \Tr[\gamma^\mu \slashed{\mathcal{B}}_c \gamma^\nu \slashed{\mathcal{B}}_b \gamma^\alpha \slashed{\mathcal{A}}_a], \nonumber \\
\mathcal{T}_{\mathcal{A} 2} ^{\mu\nu\alpha} + \mathcal{T}_{\mathcal{B} 2} ^{\mu\nu\alpha} & = & m_f ^2 \Big(\Tr[\gamma^\mu e_1 \gamma^\alpha e_2 \gamma^\nu \slashed{\mathcal{A}}_c] + \Tr[\gamma^\mu \slashed{\mathcal{B}}_c \gamma^\nu e_2 \gamma^\alpha e_1] \Big), \nonumber \\
\mathcal{T}_{\mathcal{A} 3} ^{\mu\nu\alpha} + \mathcal{T}_{\mathcal{B} 3} ^{\mu\nu\alpha} & = & m_f ^2 \Big( \Tr[\gamma^\mu e_1 \gamma^\alpha \slashed{\mathcal{A}}_b \gamma^\nu e_3] + \Tr[\gamma^\mu e_3 \gamma^\nu \slashed{\mathcal{B}}_b \gamma^\alpha e_1] \Big), \nonumber \\
\mathcal{T}_{\mathcal{A} 4} ^{\mu\nu\alpha} + \mathcal{T}_{\mathcal{B} 4} ^{\mu\nu\alpha} & = & m_f ^2 \Big( \Tr[\gamma^\mu \slashed{\mathcal{A}}_a \gamma^\alpha e_2 \gamma^\nu e_3] + \Tr[\gamma^\mu e_3 \gamma^\nu e_2 \gamma^\alpha \slashed{\mathcal{B}}_a] \Big), \nonumber \\
\mathcal{T}_{\mathcal{A} 5} ^{\mu\nu\alpha} + \mathcal{T}_{\mathcal{B} 5} ^{\mu\nu\alpha} & = & -\frac{i}{s} \Big( \Tr[\gamma^\mu \slashed{\mathcal{A}}_a \gamma^\alpha e_2 \gamma_\perp ^\nu e_3] + \Tr[\gamma^\mu e_3 \gamma_\perp ^\nu e_2 \gamma^\alpha \slashed{\mathcal{B}}_a]\Big), \nonumber \\
\mathcal{T}_{\mathcal{A} 6} ^{\mu\nu\alpha} + \mathcal{T}_{\mathcal{B} 6} ^{\mu\nu\alpha} & = & -\frac{i}{s}\Big( \Tr[\gamma_\perp ^\mu e_1 \gamma^\alpha \slashed{\mathcal{A}}_b \gamma^\nu e_3 ] + \Tr[\gamma_\perp ^\mu e_3 \gamma^\nu \slashed{\mathcal{B}}_b \gamma^\alpha e_1] \Big), \nonumber \\
\mathcal{T}_{\mathcal{A} 7} ^{\mu\nu\alpha} + \mathcal{T}_{\mathcal{B} 7} ^{\mu\nu\alpha} & = & -\frac{i}{s}\Big( \Tr[\gamma^\mu e_1 \gamma_\perp ^\alpha e_2 \gamma^\nu \slashed{\mathcal{A}}_c] + \Tr[\gamma^\mu \slashed{\mathcal{B}}_c \gamma^\nu e_2 \gamma_\perp ^\alpha e_1] \Big), \nonumber \\
\mathcal{T}_{\mathcal{A} 8} ^{\mu\nu\alpha} + \mathcal{T}_{\mathcal{B} 8} ^{\mu\nu\alpha} & = & \frac{i q_f B}{t_1 t_2 t_3 - t_1 - t_2 - t_3} \Big( \Tr[\gamma^\mu \slashed{\mathcal{A}}_a \gamma^\alpha \gamma_\parallel ^\nu] + \Tr[\gamma^\mu \gamma_\parallel ^\nu \gamma^\alpha \slashed{\mathcal{B}}_a] \Big), \nonumber \\
\mathcal{T}_{\mathcal{A} 9} ^{\mu\nu\alpha} + \mathcal{T}_{\mathcal{B} 9} ^{\mu\nu\alpha} & = & \frac{i q_f B}{t_1 t_2 t_3 - t_1 - t_2 - t_3} \left( \Tr[\gamma_\parallel ^\mu \gamma^\alpha \slashed{\mathcal{A}}_b \gamma^\nu] + \Tr[\gamma_\parallel ^\mu \gamma^\nu \slashed{\mathcal{B}}_b \gamma^\alpha] \right), \nonumber \\
\mathcal{T}_{\mathcal{A} 10} ^{\mu\nu\alpha} + \mathcal{T}_{\mathcal{B} 10} ^{\mu\nu\alpha} & = & \frac{i q_f B}{t_1 t_2 t_3 - t_1 - t_2 - t_3} \left( \Tr[\gamma^\mu \gamma_\parallel ^\alpha \gamma^\nu \slashed{\mathcal{A}}_c] + \Tr[\gamma^\mu \slashed{\mathcal{B}}_c \gamma^\nu \gamma_\parallel ^\alpha] \right), \nonumber \\
\mathcal{T}_{\mathcal{A} 11} ^{\mu\nu\alpha} + \mathcal{T}_{\mathcal{B} 11} ^{\mu\nu\alpha} & = & \frac{i q_f B t_1}{2\left( t_1 t_2 t_3 - t_1 - t_2 - t_3\right)} \left( \Tr[\gamma^\mu \slashed{\mathcal{A}}_a \gamma^\alpha \gamma^\beta \gamma^\nu \gamma^\sigma] \hat{F}_{\beta\sigma} + \Tr[\gamma^\mu \gamma^\beta \gamma^\nu \gamma^\sigma \gamma^\alpha \slashed{\mathcal{B}}_a] \hat{F}_{\beta\sigma} \right), \nonumber \\
\mathcal{T}_{\mathcal{A} 12} ^{\mu\nu\alpha} + \mathcal{T}_{\mathcal{B} 12} ^{\mu\nu\alpha} & = & -\frac{i q_f B} t_2{t_1 t_2 t_3 - t_1 - t_2 - t_3} \left( \Tr[\gamma^\mu \gamma^\beta \gamma^\alpha \slashed{\mathcal{A}}_b \gamma^\nu \gamma^\sigma] \hat{F}_{\beta\sigma} + \Tr[\gamma^\mu \gamma^\beta \gamma^\nu \slashed{\mathcal{B}}_b \gamma^\alpha \gamma^\sigma] \hat{F}_{\beta\sigma} \right), \nonumber \\
\mathcal{T}_{\mathcal{A} 13} ^{\mu\nu\alpha} + \mathcal{T}_{\mathcal{B} 13} ^{\mu\nu\alpha} & = & \frac{i q_f B t_3}{t_1 t_2 t_3 - t_1 - t_2 - t_3} \left( \Tr[\gamma^\mu \gamma^\beta \gamma^\alpha \gamma^\sigma \gamma^\nu \slashed{\mathcal{A}}_c] \hat{F}_{\beta\sigma} + \Tr[\gamma^\mu \slashed{\mathcal{B}}_c \gamma^\nu \gamma^\beta \gamma^\alpha \gamma^\sigma] \hat{F}_{\beta\sigma} \right),
\ee
where
\be
\slashed{\mathcal{A}}_a & = & - \frac{s_3 \omega_{p_1} + s_2 \omega_q}{s \omega_q} \slashed{q}_\parallel e_1 + \frac{1}{\omega_q} \frac{\left( t_3 \omega_{p_1} + t_2 \omega_q \right) \slashed{q}_\perp - t_2 t_3 \omega_{p_2} \gamma^\sigma\hat{F}_{\sigma_\beta} q_\perp ^\beta}{t_1 t_2 t_3 - t_1 - t_2 - t_3}, \nonumber \\
\slashed{\mathcal{A}}_b & = & \frac{s_1 \omega_q + s_3 \omega_{p_2}}{s \omega_q} \slashed{q}_\parallel e_2 + \frac{1}{\omega_{q}} \frac{\left( -t_1 \omega_q - t_3 \omega_{p_2}\right) \slashed{q}_\perp - t_1 t_3 \omega_{p_1} \gamma^\sigma \hat{F}_{\sigma\beta} q_\perp ^\beta }{t_1 t_2 t_3 - t_1 - t_2 - t_3}, \nonumber \\
\slashed{\mathcal{A}}_c & = & \frac{s_1 \omega_{p_1} - s_2 \omega_{p_2}}{s \omega_q} \slashed{q}_\parallel e_3 + \frac{1}{\omega_{q}} \frac{\left( -t_1 \omega_{p_1} + t_2 \omega_{p_2}\right) \slashed{q}_\perp + t_1 t_2 \omega_{q} \gamma^\sigma \hat{F}_{\sigma\beta} q_\perp ^\beta }{t_1 t_2 t_3 - t_1 - t_2 - t_3}, \nonumber \\
\slashed{\mathcal{B}}_a & = & \frac{s_3 \omega_{p_1} + s_2 \omega_q}{s \omega_q} \slashed{q}_\parallel e_1 + \frac{1}{\omega_q} \frac{\left( -t_3 \omega_{p_1} - t_2 \omega_q \right) \slashed{q}_\perp - t_2 t_3 \omega_{p_2} \gamma^\sigma\hat{F}_{\sigma_\beta} q_\perp ^\beta}{t_1 t_2 t_3 - t_1 - t_2 - t_3}, \nonumber \\
\slashed{\mathcal{B}}_b & = & -\frac{s_1 \omega_q + s_3 \omega_{p_2}}{s \omega_q} \slashed{q}_\parallel e_2 + \frac{1}{\omega_{q}} \frac{\left( t_1 \omega_q + t_3 \omega_{p_2}\right) \slashed{q}_\perp - t_1 t_3 \omega_{p_1} \gamma^\sigma \hat{F}_{\sigma\beta} q_\perp ^\beta }{t_1 t_2 t_3 - t_1 - t_2 - t_3}, \nonumber \\
\slashed{\mathcal{B}}_c & = & -\frac{s_1 \omega_{p_1} - s_2 \omega_{p_2}}{s \omega_q} \slashed{q}_\parallel e_3 + \frac{1}{\omega_{q}} \frac{\left( t_1 \omega_{p_1} - t_2 \omega_{p_2}\right) \slashed{q}_\perp + t_1 t_2 \omega_{q} \gamma^\sigma \hat{F}_{\sigma\beta} q_\perp ^\beta }{t_1 t_2 t_3 - t_1 - t_2 - t_3},
\ee

\section{Explicit expression for the functions in the vertex}\label{Appendix:functions}

The functions that define the coefficients of the basis are given by
\be
\mathcal{F} & = & \frac{1}{q_f B}\bigg( \csch(\tau)  \cosh (\tau  v_1) \cosh (\tau  v_2) \cosh (\tau  (v_1+v_2-1)) \left(\tanh (\tau  v_1) \left(\omega_q^2 \tanh (\tau  v_2)-(\omega_p-\omega_q)^2 \tanh (\tau  (v_1+v_2-1))\right) \right.\nonumber\\
& & \left.-\omega_p^2 \tanh (\tau  v_2) \tanh (\tau  (v_1+v_2-1))\right)+\tau  v_2 \omega_p^2 (v_1+v_2-1)+\tau  v_1 (v_1+v_2-1) (\omega_p-\omega_q)^2-\tau  v_1 v_2 \omega_q^2 \bigg),
\ee

\be
\mathcal{G}_1 & = & \left(q_f B + m_f ^2 \tau \right) \bigg( 3 \text{sign}(q_f B) (\text{sign}(q_f B)+1) \omega_p \cosh (2 \tau  v_1)-\text{sign}(q_f B) \omega_p \cosh (2 \tau  (2 v_1+v_2-1)) \nonumber\\
& & +\text{sign}(q_f B) \omega_p \cosh (2 \tau  (v_1+2 v_2-1))-\text{sign}(q_f B) \omega_q \cosh (2 \tau  (v_1-v_2))-3 \text{sign}(q_f B) \omega_q \cosh (2 \tau  (v_1+v_2-1)) \nonumber\\
& & +\text{sign}(q_f B) \omega_q \cosh (2 \tau  (2 v_1+v_2-1))-3 \text{sign}(q_f B) \omega_p \cosh (2 \tau  v_2)+3 \text{sign}(q_f B) \omega_q \cosh (2 \tau  v_2)+\omega_p \cosh (2 \tau  (v_1-1))\nonumber \\
& & -2 \omega_p \cosh (2 \tau  (2 v_1+v_2-1))+2 \omega_p \cosh (2 \tau  (v_1+2 v_2-1))-2 \omega_q \cosh (2 \tau  (v_1-v_2))-3 \omega_q \cosh (2 \tau  (v_1+v_2-1)) \nonumber\\
& & -\omega_q \cosh (2 \tau  (v_1+v_2))+2 \omega_q \cosh (2 \tau  (2 v_1+v_2-1))-\omega_p \cosh (2 \tau  (v_2-1))-3 \omega_p \cosh (2 \tau  v_2) \nonumber\\
& & +\omega_q \cosh (2 \tau  (v_2-1))+3 \omega_q \cosh (2 \tau  v_2)\bigg),
\ee

\be
\mathcal{G}_2 & = & \text{sign}(q_f B) \Big(4 q_f B \omega_p \sinh ^2(\tau  v_1) \cosh ^2(\tau  v_2)-4 q_f B \omega_p \cosh ^2(\tau  v_1) \sinh ^2(\tau  v_2) \nonumber\\
& & -4 q_f B \tau  \omega_p \csch(\tau ) \sinh (\tau  v_1) \cosh (\tau  v_2) \sinh (\tau  (v_1+v_2-1))+4 q_f B \tau  \omega_p \csch(\tau ) \cosh (\tau  v_1) \sinh (\tau  v_2) \sinh (\tau  (v_1+v_2-1)) \nonumber\\
& & +q_f B \omega_q \sinh (2 \tau  v_1) \sinh (2 \tau  v_2)-4 q_f B \omega_q \sinh ^2(\tau  v_1) \cosh ^2(\tau  v_2)-8 q_f B \tau  \omega_q \csch(\tau ) \sinh (\tau  v_1) \sinh (\tau  v_2) \cosh (\tau  (v_1+v_2-1)) \nonumber\\
& & +4 q_f B \tau  \omega_q \csch(\tau ) \sinh (\tau  v_1) \cosh (\tau  v_2) \sinh (\tau  (v_1+v_2-1))-3 m_f ^2 \tau  \omega_q \nonumber\\
& & +4 m_f ^2 \tau  \omega_p \sinh ^2(\tau  v_1) \cosh ^2(\tau  v_2)-4 m_f ^2 \tau  \omega_p \cosh ^2(\tau  v_1) \sinh ^2(\tau  v_2)+m_f ^2 \tau  \omega_q \sinh (2 \tau  v_1) \sinh (2 \tau  v_2) \nonumber\\
& & -m_f ^2 \tau  \omega_q \cosh (2 \tau  (v_1+v_2-1))-4 m_f ^2 \tau  \omega_q \sinh ^2(\tau  v_1) \cosh ^2(\tau  v_2)+m_f ^2 \tau  (\omega_p+\omega_q) \cosh (2 \tau  (v_2-1))+m_f ^2 \tau  \omega_p \cosh (2 \tau  v_2)\Big) \nonumber\\
& & -4 q_f B \tau  \omega_p \csch(\tau ) \sinh (\tau  v_1) \cosh (\tau  v_2) \sinh (\tau  (v_1+v_2-1))+12 q_f B \omega_q \sinh ^2(\tau  v_1) \sinh ^2(\tau  v_2) \nonumber\\
& & -q_f B \omega_q \sinh (2 \tau  v_1) \sinh (2 \tau  v_2)+4 q_f B \tau  \omega_q \csch(\tau ) \sinh (\tau  v_1) \cosh (\tau  v_2) \sinh (\tau  (v_1+v_2-1)) \nonumber\\
& & -m_f ^2 \text{sign}(q_f B) \tau  (\omega_p-2 \omega_q) \cosh (2 \tau  (v_1-1))-m_f ^2 (\text{sign}(q_f B)-1) \tau  (\omega_p-\omega_q) \cosh (2 \tau  v_1)+m_f ^2 \tau  \omega_q \nonumber\\
& & -m_f ^2 \tau  \omega_p \cosh (2 \tau  (2 v_1+v_2-1))+m_f ^2 \tau  \omega_p \cosh (2 \tau  (v_1+2 v_2-1))+12 m_f ^2 \tau  \omega_q \sinh ^2(\tau  v_1) \sinh ^2(\tau  v_2) \nonumber\\
& & -m_f ^2 \tau  \omega_q \sinh (2 \tau  v_1) \sinh (2 \tau  v_2)-m_f ^2 \tau  \omega_q \cosh (2 \tau  (v_1+v_2-1)) +m_f ^2 \tau  \omega_q \cosh (2 \tau  (2 v_1+v_2-1))\nonumber\\
& & -m_f ^2 \tau  \omega_p \cosh (2 \tau  v_2),
\ee

\be
\mathcal{G}_{10} & = & \text{sign}(q_f B) \Big(4 q_f B (\omega_p-\omega_q) \cosh ^2(\tau  v_2) \sinh ^2(\tau  (v_1+v_2-1)) \nonumber\\
& & +4 q_f B \tau  \csch(\tau ) (\omega_p-\omega_q) \sinh (\tau  v_1) \cosh (\tau  v_2) \sinh (\tau  (v_1+v_2-1))-q_f B \omega_p \sinh (2 \tau  v_1) \sinh (2 \tau  (v_1+v_2-1)) \nonumber\\
& & +q_f B \omega_p \sinh (2 \tau  v_2) \sinh (2 \tau  (v_1+v_2-1))-4 q_f B \omega_p \cosh ^2(\tau  v_1) \sinh ^2(\tau  (v_1+v_2-1)) \nonumber\\
& & -4 q_f B \tau  \omega_p \csch(\tau ) \cosh (\tau  v_1) \sinh (\tau  v_2) \sinh (\tau  (v_1+v_2-1))+q_f B \omega_q \sinh (2 \tau  v_1) \sinh (2 \tau  (v_1+v_2-1)) \nonumber\\
& & +2 q_f B \omega_q (\cosh (2 \tau  v_1)+\cosh (2 \tau  v_2)-2) \cosh ^2(\tau  (v_1+v_2-1))+8 q_f B \tau  \omega_q \csch(\tau ) \sinh (\tau  v_1) \sinh (\tau  v_2) \cosh (\tau  (v_1+v_2-1)) \nonumber\\
& & -4 m_f ^2 \tau  \omega_p \cosh (2 \tau  v_1) \sinh ^2(\tau  (v_1+v_2-1))+4 m_f ^2 \tau  \omega_p \cosh (2 \tau  v_2) \sinh ^2(\tau  (v_1+v_2-1)) \nonumber\\
& & -4 m_f ^2 \tau  \omega_q \sinh ^2(\tau  (v_1+v_2-1))+2 m_f ^2 \tau  \omega_q \sinh (2 \tau  v_1) \sinh (2 \tau  (v_1+v_2-1))+2 m_f ^2 \tau  \omega_q \sinh (2 \tau  v_2) \sinh (2 \tau  (v_1+v_2-1)) \nonumber\\
& & -4 m_f ^2 \tau  \omega_q \cosh (2 \tau  v_2) \sinh ^2(\tau  (v_1+v_2-1))\Big)+12 q_f B \omega_p \sinh ^2(\tau  v_1) \sinh ^2(\tau  (v_1+v_2-1)) \nonumber\\
& & -12 q_f B \omega_p \sinh ^2(\tau  v_2) \sinh ^2(\tau  (v_1+v_2-1))+q_f B \omega_p \sinh (2 \tau  v_1) \sinh (2 \tau  (v_1+v_2-1)) \nonumber\\
& & -q_f B \omega_p \sinh (2 \tau  v_2) \sinh (2 \tau  (v_1+v_2-1))-4 q_f B \tau  \omega_p \csch(\tau ) \sinh (\tau  v_1) \cosh (\tau  v_2) \sinh (\tau  (v_1+v_2-1)) \nonumber\\
& & -12 q_f B \omega_q \sinh ^2(\tau  v_1) \sinh ^2(\tau  (v_1+v_2-1))-q_f B \omega_q \sinh (2 \tau  v_1) \sinh (2 \tau  (v_1+v_2-1)) \nonumber\\
& & +4 q_f B \tau  \omega_q \csch(\tau ) \sinh (\tau  v_1) \cosh (\tau  v_2) \sinh (\tau  (v_1+v_2-1))-2 m_f ^2 \tau  \omega_q+2 m_f ^2 \tau  \omega_q \cosh (2 \tau  v_1) \nonumber\\
& & +4 m_f ^2 \tau  \omega_p \cosh (2 \tau  v_1) \sinh ^2(\tau  (v_1+v_2-1))-4 m_f ^2 \tau  \omega_p \cosh (2 \tau  v_2) \sinh ^2(\tau  (v_1+v_2-1))+4 m_f ^2 \tau  \omega_q \sinh ^2(\tau  (v_1+v_2-1)) \nonumber\\
& & -2 m_f ^2 \tau  \omega_q \cosh (2 \tau  (v_1-v_2))-4 m_f ^2 \tau  \omega_q \cosh (2 \tau  v_1) \sinh ^2(\tau  (v_1+v_2-1))+2 m_f ^2 \tau  \omega_q \cosh (2 \tau  v_2),
\ee

\be
\mathcal{H}_1 & = & -3 v_1^2 \cosh (2 (v_2-1) \tau ) \omega_p^3-3 v_2^2 \cosh (2 (v_2-1) \tau ) \omega_p^3+3 v_1 \cosh (2 (v_2-1) \tau ) \omega_p^3 \nonumber\\
& & -6 v_1 v_2 \cosh (2 (v_2-1) \tau ) \omega_p^3+3 v_2 \cosh (2 (v_2-1) \tau ) \omega_p^3-6 v_1^2 \cosh (2 v_2 \tau ) \omega_p^3-6 v_2^2 \cosh (2 v_2 \tau ) \omega_p^3+6 v_1 \cosh (2 v_2 \tau ) \omega_p^3\nonumber\\
& & -12 v_1 v_2 \cosh (2 v_2 \tau ) \omega_p^3+6 v_2 \cosh (2 v_2 \tau ) \omega_p^3-3 v_1^2 \cosh (2 (2 v_1+v_2-1) \tau ) \omega_p^3-3 v_2^2 \cosh (2 (2 v_1+v_2-1) \tau ) \omega_p^3\nonumber\\
& & +3 v_1 \cosh (2 (2 v_1+v_2-1) \tau ) \omega_p^3-6 v_1 v_2 \cosh (2 (2 v_1+v_2-1) \tau ) \omega_p^3+3 v_2 \cosh (2 (2 v_1+v_2-1) \tau ) \omega_p^3 \nonumber\\
& & +3 v_1^2 \cosh (2 (v_1+2 v_2-1) \tau ) \omega_p^3+3 v_2^2 \cosh (2 (v_1+2 v_2-1) \tau ) \omega_p^3-3 v_1 \cosh (2 (v_1+2 v_2-1) \tau ) \omega_p^3\nonumber\\
& & +6 v_1 v_2 \cosh (2 (v_1+2 v_2-1) \tau ) \omega_p^3-3 v_2 \cosh (2 (v_1+2 v_2-1) \tau ) \omega_p^3-6 \omega_q \omega_p^2-3 v_1^2 \omega_q \cosh (2 (v_1-v_2) \tau ) \omega_p^2 \nonumber\\
& & -3 v_2^2 \omega_q \cosh (2 (v_1-v_2) \tau ) \omega_p^2+6 v_1 \omega_q \cosh (2 (v_1-v_2) \tau ) \omega_p^2-6 v_1 v_2 \omega_q \cosh (2 (v_1-v_2) \tau ) \omega_p^2+6 v_2 \omega_q \cosh (2 (v_1-v_2) \tau ) \omega_p^2 \nonumber\\
& & -3 \omega_q \cosh (2 (v_1-v_2) \tau ) \omega_p^2+9 v_1^2 \omega_q \cosh (2 (v_2-1) \tau ) \omega_p^2+3 v_2^2 \omega_q \cosh (2 (v_2-1) \tau ) \omega_p^2-9 v_1 \omega_q \cosh (2 (v_2-1) \tau ) \omega_p^2 \nonumber \\
& & +12 v_1 v_2 \omega_q \cosh (2 (v_2-1) \tau ) \omega_p^2-6 v_2 \omega_q \cosh (2 (v_2-1) \tau ) \omega_p^2+18 v_1^2 \omega_q \cosh (2 v_2 \tau ) \omega_p^2+6 v_2^2 \omega_q \cosh (2 v_2 \tau ) \omega_p^2 \nonumber\\
& & -18 v_1 \omega_q \cosh (2 v_2 \tau ) \omega_p^2+24 v_1 v_2 \omega_q \cosh (2 v_2 \tau ) \omega_p^2-12 v_2 \omega_q \cosh (2 v_2 \tau ) \omega_p^2+6 \omega_q \cosh (2 v_2 \tau ) \omega_p^2 \nonumber\\
& & -6 v_1^2 \omega_q \cosh (2 (v_1+v_2-1) \tau ) \omega_p^2-6 v_2^2 \omega_q \cosh (2 (v_1+v_2-1) \tau ) \omega_p^2+12 v_1 \omega_q \cosh (2 (v_1+v_2-1) \tau ) \omega_p^2 \nonumber\\
& & -12 v_1 v_2 \omega_q \cosh (2 (v_1+v_2-1) \tau ) \omega_p^2+12 v_2 \omega_q \cosh (2 (v_1+v_2-1) \tau ) \omega_p^2-3 v_1^2 \omega_q \cosh (2 (v_1+v_2) \tau ) \omega_p^2 \nonumber\\
& & -3 v_2^2 \omega_q \cosh (2 (v_1+v_2) \tau ) \omega_p^2+6 v_1 \omega_q \cosh (2 (v_1+v_2) \tau ) \omega_p^2-6 v_1 v_2 \omega_q \cosh (2 (v_1+v_2) \tau ) \omega_p^2+6 v_2 \omega_q \cosh (2 (v_1+v_2) \tau ) \omega_p^2 \nonumber\\
& & -3 \omega_q \cosh (2 (v_1+v_2) \tau ) \omega_p^2+9 v_1^2 \omega_q \cosh (2 (2 v_1+v_2-1) \tau ) \omega_p^2+3 v_2^2 \omega_q \cosh (2 (2 v_1+v_2-1) \tau ) \omega_p^2 \nonumber\\
& & -9 v_1 \omega_q \cosh (2 (2 v_1+v_2-1) \tau ) \omega_p^2+12 v_1 v_2 \omega_q \cosh (2 (2 v_1+v_2-1) \tau ) \omega_p^2-6 v_2 \omega_q \cosh (2 (2 v_1+v_2-1) \tau ) \omega_p^2 \nonumber\\
& & -6 v_1^2 \omega_q \cosh (2 (v_1+2 v_2-1) \tau ) \omega_p^2+3 v_1 \omega_q \cosh (2 (v_1+2 v_2-1) \tau ) \omega_p^2-6 v_1 v_2 \omega_q \cosh (2 (v_1+2 v_2-1) \tau ) \omega_p^2 \nonumber\\
& & +6 \omega_q^2 \omega_p+3 \left((v_2-1) v_2 \omega_p^2+v_1 (2 v_2-1) (\omega_p-\omega_q) \omega_p+v_1^2 (\omega_p-\omega_q)^2\right) \cosh (2 (v_1-1) \tau ) \omega_p \nonumber\\
& & +2 ((v_1+v_2-1) \omega_p-v_1 \omega_q+\omega_q) (3 v_1 \omega_p+3 v_2 \omega_p-3 v_1 \omega_q-3 \omega_q \nonumber\\
& & +\text{sign}(q_f B) ((3 v_1+3 v_2-2) \omega_p-3 v_1 \omega_q+\omega_q)) \cosh (2 v_1 \tau ) \omega_p+6 v_1^2 \omega_q^2 \cosh (2 (v_1-v_2) \tau ) \omega_p \nonumber\\
& & -9 v_1 \omega_q^2 \cosh (2 (v_1-v_2) \tau ) \omega_p+6 v_1 v_2 \omega_q^2 \cosh (2 (v_1-v_2) \tau ) \omega_p-3 v_2 \omega_q^2 \cosh (2 (v_1-v_2) \tau ) \omega_p+3 \omega_q^2 \cosh (2 (v_1-v_2) \tau ) \omega_p \nonumber\\
& & -9 v_1^2 \omega_q^2 \cosh (2 (v_2-1) \tau ) \omega_p+9 v_1 \omega_q^2 \cosh (2 (v_2-1) \tau ) \omega_p-6 v_1 v_2 \omega_q^2 \cosh (2 (v_2-1) \tau ) \omega_p+3 v_2 \omega_q^2 \cosh (2 (v_2-1) \tau ) \omega_p \nonumber\\
& & -18 v_1^2 \omega_q^2 \cosh (2 v_2 \tau ) \omega_p+18 v_1 \omega_q^2 \cosh (2 v_2 \tau ) \omega_p-12 v_1 v_2 \omega_q^2 \cosh (2 v_2 \tau ) \omega_p+6 v_2 \omega_q^2 \cosh (2 v_2 \tau ) \omega_p-6 \omega_q^2 \cosh (2 v_2 \tau ) \omega_p \nonumber\\
& & +12 v_1^2 \omega_q^2 \cosh (2 (v_1+v_2-1) \tau ) \omega_p-18 v_1 \omega_q^2 \cosh (2 (v_1+v_2-1) \tau ) \omega_p+12 v_1 v_2 \omega_q^2 \cosh (2 (v_1+v_2-1) \tau ) \omega_p \nonumber\\
& & -6 v_2 \omega_q^2 \cosh (2 (v_1+v_2-1) \tau ) \omega_p+6 v_1^2 \omega_q^2 \cosh (2 (v_1+v_2) \tau ) \omega_p-9 v_1 \omega_q^2 \cosh (2 (v_1+v_2) \tau ) \omega_p \nonumber\\
& & +6 v_1 v_2 \omega_q^2 \cosh (2 (v_1+v_2) \tau ) \omega_p-3 v_2 \omega_q^2 \cosh (2 (v_1+v_2) \tau ) \omega_p+3 \omega_q^2 \cosh (2 (v_1+v_2) \tau ) \omega_p \nonumber\\
& & -9 v_1^2 \omega_q^2 \cosh (2 (2 v_1+v_2-1) \tau ) \omega_p+9 v_1 \omega_q^2 \cosh (2 (2 v_1+v_2-1) \tau ) \omega_p-6 v_1 v_2 \omega_q^2 \cosh (2 (2 v_1+v_2-1) \tau ) \omega_p \nonumber\\
& & +3 v_2 \omega_q^2 \cosh (2 (2 v_1+v_2-1) \tau ) \omega_p+3 v_1^2 \omega_q^2 \cosh (2 (v_1+2 v_2-1) \tau ) \omega_p-3 v_1^2 \omega_q^3 \cosh (2 (v_1-v_2) \tau )+3 v_1 \omega_q^3 \cosh (2 (v_1-v_2) \tau ) \nonumber\\
& & +3 v_1^2 \omega_q^3 \cosh (2 (v_2-1) \tau )-3 v_1 \omega_q^3 \cosh (2 (v_2-1) \tau )+6 v_1^2 \omega_q^3 \cosh (2 v_2 \tau )-6 v_1 \omega_q^3 \cosh (2 v_2 \tau ) \nonumber\\
& & -6 v_1^2 \omega_q^3 \cosh (2 (v_1+v_2-1) \tau )+6 v_1 \omega_q^3 \cosh (2 (v_1+v_2-1) \tau )-3 v_1^2 \omega_q^3 \cosh (2 (v_1+v_2) \tau )+3 v_1 \omega_q^3 \cosh (2 (v_1+v_2) \tau ) \nonumber\\
& & +3 v_1^2 \omega_q^3 \cosh (2 (2 v_1+v_2-1) \tau )-3 v_1 \omega_q^3 \cosh (2 (2 v_1+v_2-1) \tau ) \nonumber\\
& & -2 \text{sign}(q_f B) \left(v_1^2 \cosh (2 (2 v_1+v_2-1) \tau ) \omega_p^3+v_2^2 \cosh (2 (2 v_1+v_2-1) \tau ) \omega_p^3-v_1 \cosh (2 (2 v_1+v_2-1) \tau ) \omega_p^3 \right. \nonumber\\
& & +2 v_1 v_2 \cosh (2 (2 v_1+v_2-1) \tau ) \omega_p^3-v_2 \cosh (2 (2 v_1+v_2-1) \tau ) \omega_p^3-v_1^2 \cosh (2 (v_1+2 v_2-1) \tau ) \omega_p^3 \nonumber\\
& & -v_2^2 \cosh (2 (v_1+2 v_2-1) \tau ) \omega_p^3+v_1 \cosh (2 (v_1+2 v_2-1) \tau ) \omega_p^3-2 v_1 v_2 \cosh (2 (v_1+2 v_2-1) \tau ) \omega_p^3 \nonumber\\
& & +v_2 \cosh (2 (v_1+2 v_2-1) \tau ) \omega_p^3+3 v_1^2 \omega_q \cosh (2 (v_1+v_2-1) \tau ) \omega_p^2+3 v_2^2 \omega_q \cosh (2 (v_1+v_2-1) \tau ) \omega_p^2 \nonumber\\
& & -2 v_1 \omega_q \cosh (2 (v_1+v_2-1) \tau ) \omega_p^2+6 v_1 v_2 \omega_q \cosh (2 (v_1+v_2-1) \tau ) \omega_p^2-2 v_2 \omega_q \cosh (2 (v_1+v_2-1) \tau ) \omega_p^2 \nonumber\\
& & -3 v_1^2 \omega_q \cosh (2 (2 v_1+v_2-1) \tau ) \omega_p^2-v_2^2 \omega_q \cosh (2 (2 v_1+v_2-1) \tau ) \omega_p^2+3 v_1 \omega_q \cosh (2 (2 v_1+v_2-1) \tau ) \omega_p^2 \nonumber\\
& & -4 v_1 v_2 \omega_q \cosh (2 (2 v_1+v_2-1) \tau ) \omega_p^2+2 v_2 \omega_q \cosh (2 (2 v_1+v_2-1) \tau ) \omega_p^2+2 v_1^2 \omega_q \cosh (2 (v_1+2 v_2-1) \tau ) \omega_p^2 \nonumber\\
& & -v_1 \omega_q \cosh (2 (v_1+2 v_2-1) \tau ) \omega_p^2+2 v_1 v_2 \omega_q \cosh (2 (v_1+2 v_2-1) \tau ) \omega_p^2+\omega_q^2 \omega_p-6 v_1^2 \omega_q^2 \cosh (2 (v_1+v_2-1) \tau ) \omega_p \nonumber\\
& & +3 v_1 \omega_q^2 \cosh (2 (v_1+v_2-1) \tau ) \omega_p-6 v_1 v_2 \omega_q^2 \cosh (2 (v_1+v_2-1) \tau ) \omega_p+v_2 \omega_q^2 \cosh (2 (v_1+v_2-1) \tau ) \omega_p \nonumber\\
& & +3 v_1^2 \omega_q^2 \cosh (2 (2 v_1+v_2-1) \tau ) \omega_p-3 v_1 \omega_q^2 \cosh (2 (2 v_1+v_2-1) \tau ) \omega_p+2 v_1 v_2 \omega_q^2 \cosh (2 (2 v_1+v_2-1) \tau ) \omega_p \nonumber
\ee
\be
\phantom{H_1 = }& & -v_2 \omega_q^2 \cosh (2 (2 v_1+v_2-1) \tau ) \omega_p-v_1^2 \omega_q^2 \cosh (2 (v_1+2 v_2-1) \tau ) \omega_p-\omega_p^2 \omega_q \nonumber\\
& & +\omega_q \left((v_1+v_2-1)^2 \omega_p^2-(2 v_1-1) (v_1+v_2-1) \omega_q \omega_p+(v_1-1) v_1 \omega_q^2\right) \cosh (2 (v_1-v_2) \tau ) \nonumber\\
& & +(\omega_p-\omega_q) \left(\left(3 v_1^2+(6 v_2-5) v_1+3 v_2^2-5 v_2+2\right) \omega_p^2-\left(6 v_1^2+6 (v_2-1) v_1-v_2+1\right) \omega_q \omega_p+v_1 (3 v_1-1) \omega_q^2\right) \cosh (2 v_2 \tau ) \nonumber\\
& &  +3 v_1^2 \omega_q^3 \cosh (2 (v_1+v_2-1) \tau )-v_1 \omega_q^3 \cosh (2 (v_1+v_2-1) \tau )-v_1^2 \omega_q^3 \cosh (2 (2 v_1+v_2-1) \tau ) \nonumber\\
& & \left. +v_1 \omega_q^3 \cosh (2 (2 v_1+v_2-1) \tau )\right),
\ee

\be
\mathcal{H}_2 & = & -v_1^2 \cosh (2 (v_2-1) \tau ) \omega_p^3-v_2^2 \cosh (2 (v_2-1) \tau ) \omega_p^3+v_1 \cosh (2 (v_2-1) \tau ) \omega_p^3 \nonumber\\
& & -2 v_1 v_2 \cosh (2 (v_2-1) \tau ) \omega_p^3+v_2 \cosh (2 (v_2-1) \tau ) \omega_p^3-2 v_1^2 \cosh (2 v_2 \tau ) \omega_p^3-2 v_2^2 \cosh (2 v_2 \tau ) \omega_p^3+2 v_1 \cosh (2 v_2 \tau ) \omega_p^3 \nonumber\\
& & -4 v_1 v_2 \cosh (2 v_2 \tau ) \omega_p^3+2 v_2 \cosh (2 v_2 \tau ) \omega_p^3-v_1^2 \cosh (2 (2 v_1+v_2-1) \tau ) \omega_p^3-v_2^2 \cosh (2 (2 v_1+v_2-1) \tau ) \omega_p^3 \nonumber\\
& & +v_1 \cosh (2 (2 v_1+v_2-1) \tau ) \omega_p^3-2 v_1 v_2 \cosh (2 (2 v_1+v_2-1) \tau ) \omega_p^3+v_2 \cosh (2 (2 v_1+v_2-1) \tau ) \omega_p^3 \nonumber\\
& & +v_1^2 \cosh (2 (v_1+2 v_2-1) \tau ) \omega_p^3+v_2^2 \cosh (2 (v_1+2 v_2-1) \tau ) \omega_p^3-v_1 \cosh (2 (v_1+2 v_2-1) \tau ) \omega_p^3 \nonumber\\
& & +2 v_1 v_2 \cosh (2 (v_1+2 v_2-1) \tau ) \omega_p^3-v_2 \cosh (2 (v_1+2 v_2-1) \tau ) \omega_p^3+8 v_1^2 \omega_q \omega_p^2+8 v_2^2 \omega_q \omega_p^2-16 v_1 \omega_q \omega_p^2+16 v_1 v_2 \omega_q \omega_p^2 \nonumber\\
& & -16 v_2 \omega_q \omega_p^2+6 \omega_q \omega_p^2+3 v_1^2 \omega_q \cosh (2 (v_1-v_2) \tau ) \omega_p^2+3 v_2^2 \omega_q \cosh (2 (v_1-v_2) \tau ) \omega_p^2-6 v_1 \omega_q \cosh (2 (v_1-v_2) \tau ) \omega_p^2 \nonumber\\
& & +6 v_1 v_2 \omega_q \cosh (2 (v_1-v_2) \tau ) \omega_p^2-6 v_2 \omega_q \cosh (2 (v_1-v_2) \tau ) \omega_p^2+3 \omega_q \cosh (2 (v_1-v_2) \tau ) \omega_p^2 \nonumber\\
& & +3 v_1^2 \omega_q \cosh (2 (v_2-1) \tau ) \omega_p^2+v_2^2 \omega_q \cosh (2 (v_2-1) \tau ) \omega_p^2-3 v_1 \omega_q \cosh (2 (v_2-1) \tau ) \omega_p^2 \nonumber\\
& & +4 v_1 v_2 \omega_q \cosh (2 (v_2-1) \tau ) \omega_p^2-2 v_2 \omega_q \cosh (2 (v_2-1) \tau ) \omega_p^2-2 v_1^2 \omega_q \cosh (2 v_2 \tau ) \omega_p^2-6 v_2^2 \omega_q \cosh (2 v_2 \tau ) \omega_p^2 \nonumber\\
& & +10 v_1 \omega_q \cosh (2 v_2 \tau ) \omega_p^2-8 v_1 v_2 \omega_q \cosh (2 v_2 \tau ) \omega_p^2+12 v_2 \omega_q \cosh (2 v_2 \tau ) \omega_p^2-6 \omega_q \cosh (2 v_2 \tau ) \omega_p^2 \nonumber\\
& & -2 v_1^2 \omega_q \cosh (2 (v_1+v_2-1) \tau ) \omega_p^2-2 v_2^2 \omega_q \cosh (2 (v_1+v_2-1) \tau ) \omega_p^2+4 v_1 \omega_q \cosh (2 (v_1+v_2-1) \tau ) \omega_p^2 \nonumber\\
& & -4 v_1 v_2 \omega_q \cosh (2 (v_1+v_2-1) \tau ) \omega_p^2+4 v_2 \omega_q \cosh (2 (v_1+v_2-1) \tau ) \omega_p^2+3 v_1^2 \omega_q \cosh (2 (v_1+v_2) \tau ) \omega_p^2 \nonumber\\
& & +3 v_2^2 \omega_q \cosh (2 (v_1+v_2) \tau ) \omega_p^2-6 v_1 \omega_q \cosh (2 (v_1+v_2) \tau ) \omega_p^2+6 v_1 v_2 \omega_q \cosh (2 (v_1+v_2) \tau ) \omega_p^2 \nonumber\\
& & -6 v_2 \omega_q \cosh (2 (v_1+v_2) \tau ) \omega_p^2+3 \omega_q \cosh (2 (v_1+v_2) \tau ) \omega_p^2+3 v_1^2 \omega_q \cosh (2 (2 v_1+v_2-1) \tau ) \omega_p^2 \nonumber\\
& & +v_2^2 \omega_q \cosh (2 (2 v_1+v_2-1) \tau ) \omega_p^2-3 v_1 \omega_q \cosh (2 (2 v_1+v_2-1) \tau ) \omega_p^2+4 v_1 v_2 \omega_q \cosh (2 (2 v_1+v_2-1) \tau ) \omega_p^2 \nonumber\\
& & -2 v_2 \omega_q \cosh (2 (2 v_1+v_2-1) \tau ) \omega_p^2-2 v_1^2 \omega_q \cosh (2 (v_1+2 v_2-1) \tau ) \omega_p^2+v_1 \omega_q \cosh (2 (v_1+2 v_2-1) \tau ) \omega_p^2 \nonumber\\
& & -2 v_1 v_2 \omega_q \cosh (2 (v_1+2 v_2-1) \tau ) \omega_p^2-16 v_1^2 \omega_q^2 \omega_p+24 v_1 \omega_q^2 \omega_p-16 v_1 v_2 \omega_q^2 \omega_p+8 v_2 \omega_q^2 \omega_p-6 \omega_q^2 \omega_p \nonumber\\
& & -6 v_1^2 \omega_q^2 \cosh (2 (v_1-v_2) \tau ) \omega_p+9 v_1 \omega_q^2 \cosh (2 (v_1-v_2) \tau ) \omega_p-6 v_1 v_2 \omega_q^2 \cosh (2 (v_1-v_2) \tau ) \omega_p \nonumber\\
& & +3 v_2 \omega_q^2 \cosh (2 (v_1-v_2) \tau ) \omega_p-3 \omega_q^2 \cosh (2 (v_1-v_2) \tau ) \omega_p-3 v_1^2 \omega_q^2 \cosh (2 (v_2-1) \tau ) \omega_p+3 v_1 \omega_q^2 \cosh (2 (v_2-1) \tau ) \omega_p \nonumber\\
& & -2 v_1 v_2 \omega_q^2 \cosh (2 (v_2-1) \tau ) \omega_p+v_2 \omega_q^2 \cosh (2 (v_2-1) \tau ) \omega_p+10 v_1^2 \omega_q^2 \cosh (2 v_2 \tau ) \omega_p-18 v_1 \omega_q^2 \cosh (2 v_2 \tau ) \omega_p \nonumber\\
& & +12 v_1 v_2 \omega_q^2 \cosh (2 v_2 \tau ) \omega_p-6 v_2 \omega_q^2 \cosh (2 v_2 \tau ) \omega_p+6 \omega_q^2 \cosh (2 v_2 \tau ) \omega_p+4 v_1^2 \omega_q^2 \cosh (2 (v_1+v_2-1) \tau ) \omega_p \nonumber\\
& & -6 v_1 \omega_q^2 \cosh (2 (v_1+v_2-1) \tau ) \omega_p+4 v_1 v_2 \omega_q^2 \cosh (2 (v_1+v_2-1) \tau ) \omega_p-2 v_2 \omega_q^2 \cosh (2 (v_1+v_2-1) \tau ) \omega_p \nonumber\\
& & -6 v_1^2 \omega_q^2 \cosh (2 (v_1+v_2) \tau ) \omega_p+9 v_1 \omega_q^2 \cosh (2 (v_1+v_2) \tau ) \omega_p-6 v_1 v_2 \omega_q^2 \cosh (2 (v_1+v_2) \tau ) \omega_p \nonumber\\
& & +3 v_2 \omega_q^2 \cosh (2 (v_1+v_2) \tau ) \omega_p-3 \omega_q^2 \cosh (2 (v_1+v_2) \tau ) \omega_p-3 v_1^2 \omega_q^2 \cosh (2 (2 v_1+v_2-1) \tau ) \omega_p \nonumber\\
& & +3 v_1 \omega_q^2 \cosh (2 (2 v_1+v_2-1) \tau ) \omega_p-2 v_1 v_2 \omega_q^2 \cosh (2 (2 v_1+v_2-1) \tau ) \omega_p+v_2 \omega_q^2 \cosh (2 (2 v_1+v_2-1) \tau ) \omega_p \nonumber\\
& & +v_1^2 \omega_q^2 \cosh (2 (v_1+2 v_2-1) \tau ) \omega_p+8 v_1^2 \omega_q^3-8 v_1 \omega_q^3-\left((v_2-1) v_2 \omega_p^2+v_1 (2 v_2-1) (\omega_p-\omega_q) \omega_p+v_1^2 (\omega_p-\omega_q)^2\right) \nonumber\\
& & \times ((2 \text{sign}(q_f B)-1) \omega_p-4 \text{sign}(q_f B) \omega_q) \cosh (2 (v_1-1) \tau ) \nonumber\\
& & +2 ((v_1+v_2-1) \omega_p-v_1 \omega_q+\omega_q) \left(\text{sign}(q_f B) \omega_p ((v_1+v_2-2) \omega_p-v_1 \omega_q+\omega_q)+\omega_p (v_2 \omega_p-4 v_2 \omega_q+3 \omega_q) \right. \nonumber\\
& & \left. +v_1 \left(\omega_p^2-5 \omega_q \omega_p+4 \omega_q^2\right)\right) \cosh (2 v_1 \tau )+3 v_1^2 \omega_q^3 \cosh (2 (v_1-v_2) \tau )-3 v_1 \omega_q^3 \cosh (2 (v_1-v_2) \tau )+v_1^2 \omega_q^3 \cosh (2 (v_2-1) \tau ) \nonumber
\ee

\be
\phantom{H_2 = } & & -v_1 \omega_q^3 \cosh (2 (v_2-1) \tau )-6 v_1^2 \omega_q^3 \cosh (2 v_2 \tau )+6 v_1 \omega_q^3 \cosh (2 v_2 \tau )-2 v_1^2 \omega_q^3 \cosh (2 (v_1+v_2-1) \tau ) \nonumber\\
& & +2 v_1 \omega_q^3 \cosh (2 (v_1+v_2-1) \tau )-2 \text{sign}(q_f B) \left(v_1^2 \cosh (2 v_2 \tau ) \omega_p^3+v_2^2 \cosh (2 v_2 \tau ) \omega_p^3-3 v_1 \cosh (2 v_2 \tau ) \omega_p^3+2 v_1 v_2 \cosh (2 v_2 \tau ) \omega_p^3 \right. \nonumber\\
& & -3 v_2 \cosh (2 v_2 \tau ) \omega_p^3+2 \cosh (2 v_2 \tau ) \omega_p^3+2 v_1^2 \omega_q \omega_p^2+2 v_2^2 \omega_q \omega_p^2+4 v_1 v_2 \omega_q \omega_p^2-3 v_1^2 \omega_q \cosh (2 v_2 \tau ) \omega_p^2 \nonumber\\
& & -v_2^2 \omega_q \cosh (2 v_2 \tau ) \omega_p^2+7 v_1 \omega_q \cosh (2 v_2 \tau ) \omega_p^2-4 v_1 v_2 \omega_q \cosh (2 v_2 \tau ) \omega_p^2+4 v_2 \omega_q \cosh (2 v_2 \tau ) \omega_p^2-3 \omega_q \cosh (2 v_2 \tau ) \omega_p^2\nonumber\\
& & +v_1^2 \omega_q \cosh (2 (v_1+v_2-1) \tau ) \omega_p^2+v_2^2 \omega_q \cosh (2 (v_1+v_2-1) \tau ) \omega_p^2-2 v_1 \omega_q \cosh (2 (v_1+v_2-1) \tau ) \omega_p^2\nonumber\\
& & +2 v_1 v_2 \omega_q \cosh (2 (v_1+v_2-1) \tau ) \omega_p^2-2 v_2 \omega_q \cosh (2 (v_1+v_2-1) \tau ) \omega_p^2-4 v_1^2 \omega_q^2 \omega_p-4 v_1 v_2 \omega_q^2 \omega_p+\omega_q^2 \omega_p\nonumber\\
& & +3 v_1^2 \omega_q^2 \cosh (2 v_2 \tau ) \omega_p-5 v_1 \omega_q^2 \cosh (2 v_2 \tau ) \omega_p+2 v_1 v_2 \omega_q^2 \cosh (2 v_2 \tau ) \omega_p-v_2 \omega_q^2 \cosh (2 v_2 \tau ) \omega_p+\omega_q^2 \cosh (2 v_2 \tau ) \omega_p\nonumber\\
& &  -2 v_1^2 \omega_q^2 \cosh (2 (v_1+v_2-1) \tau ) \omega_p+3 v_1 \omega_q^2 \cosh (2 (v_1+v_2-1) \tau ) \omega_p-2 v_1 v_2 \omega_q^2 \cosh (2 (v_1+v_2-1) \tau ) \omega_p\nonumber\\
& & +v_2 \omega_q^2 \cosh (2 (v_1+v_2-1) \tau ) \omega_p+2 v_1^2 \omega_q^3-\omega_p^2 \omega_q\nonumber\\
& & +\omega_q \left((v_1+v_2-1)^2 \omega_p^2-(2 v_1-1) (v_1+v_2-1) \omega_q \omega_p+(v_1-1) v_1 \omega_q^2\right) \cosh (2 (v_1-v_2) \tau )\nonumber\\
& &  -(\omega_p+\omega_q) \left(v_1^2 (\omega_p-\omega_q)^2+v_1 ((2 v_2-1) \omega_p+\omega_q) (\omega_p-\omega_q)+v_2 \omega_p ((v_2-1) \omega_p+\omega_q)\right) \cosh (2 (v_2-1) \tau )\nonumber\\
& & \left. -v_1^2 \omega_q^3 \cosh (2 v_2 \tau )+v_1 \omega_q^3 \cosh (2 v_2 \tau )+v_1^2 \omega_q^3 \cosh (2 (v_1+v_2-1) \tau )-v_1 \omega_q^3 \cosh (2 (v_1+v_2-1) \tau )\right)\nonumber\\
& & +3 v_1^2 \omega_q^3 \cosh (2 (v_1+v_2) \tau )-3 v_1 \omega_q^3 \cosh (2 (v_1+v_2) \tau )+v_1^2 \omega_q^3 \cosh (2 (2 v_1+v_2-1) \tau )-v_1 \omega_q^3 \cosh (2 (2 v_1+v_2-1) \tau ),
\ee

\be
\mathcal{H}_{10} & = & 4\bigg(\sinh (\tau  (v_1+v_2-1)) (v_1 (\omega_p-\omega_q)+v_2 \omega_p) \Big((\text{sign}(q_f B) (\omega_p+\omega_q)-\omega_p+\omega_q)  \nonumber\\
& & \times \sinh (\tau  (v_1-v_2+1)) (\omega_p (v_1+v_2-1)-v_1 \omega_q+\omega_q) \nonumber\\
& & -((\text{sign}(q_f B)-1) \omega_p-2 \text{sign}(q_f B) \omega_q) \sinh (\tau  (-v_1+v_2+1)) (\omega_p (v_1+v_2-1)-v_1 \omega_q) \nonumber\\
& & +\text{sign}(q_f B) (-2 \omega_q \sinh (\tau  (v_1+v_2-1)) (v_1 (\omega_p-\omega_q)+v_2 \omega_p)-(\omega_p-\omega_q) \sinh (\tau  (3 v_1+v_2-1))  \nonumber\\
& & \times (\omega_p (v_1+v_2-1)-v_1 \omega_q+\omega_q) \nonumber\\
& & +\omega_p \sinh (\tau  (v_1+3 v_2-1)) (\omega_p (v_1+v_2-1)-v_1 \omega_q))+v_1 \omega_p^2 \sinh (\tau  (3 v_1+v_2-1))+v_2 \omega_p^2 \sinh (\tau  (3 v_1+v_2-1)) \nonumber\\
& & -\omega_p^2 \sinh (\tau  (3 v_1+v_2-1))-v_1 \omega_p^2 \sinh (\tau  (v_1+3 v_2-1))-v_2 \omega_p^2 \sinh (\tau  (v_1+3 v_2-1))+\omega_p^2 \sinh (\tau  (v_1+3 v_2-1)) \nonumber\\
& & +2 v_1 \omega_p \omega_q \sinh (\tau  (v_1+v_2-1))+2 v_2 \omega_p \omega_q \sinh (\tau  (v_1+v_2-1))-4 \omega_p \omega_q \sinh (\tau  (v_1+v_2-1)) \nonumber\\
& & -2 v_1 \omega_p \omega_q \sinh (\tau  (3 v_1+v_2-1))-v_2 \omega_p \omega_q \sinh (\tau  (3 v_1+v_2-1))+2 \omega_p \omega_q \sinh (\tau  (3 v_1+v_2-1)) \nonumber\\
& & +v_1 \omega_p \omega_q \sinh (\tau  (v_1+3 v_2-1))-2 v_1 \omega_q^2 \sinh (\tau  (v_1+v_2-1))+2 \omega_q^2 \sinh (\tau  (v_1+v_2-1))+v_1 \omega_q^2 \sinh (\tau  (3 v_1+v_2-1)) \nonumber\\
& & -\omega_q^2 \sinh (\tau  (3 v_1+v_2-1))\Big)-4 \omega_q \sinh ^2(\tau  v_1) \sinh ^2(\tau  v_2) \left(\omega_p^2 (v_1+v_2-1)^2 \right.\nonumber\\
& & \left. -(2 v_1-1) \omega_p \omega_q (v_1+v_2-1)+(v_1-1) v_1 \omega_q^2\right)\bigg),
\ee
where we have imposed the conservation of energy $\omega_q = \omega_{p_1} + \omega_{p_2}$, and so it is possible to write $\omega_{p_2} = \omega_q - \omega_{p_1}$, with $\omega_p \equiv \omega_{p_1}$.
\end{widetext}

%\clearpage
\bibliography{biblio}

\end{document}